\documentclass[aps,prx,reprint,superscriptaddress,longbibliography,floatfix]{revtex4-2}
\usepackage[T1]{fontenc}
\ifdefined\DeclareUnicodeCharacter\DeclareUnicodeCharacter{2009}{\,}\fi
\PassOptionsToPackage{stretch=20,shrink=30}{microtype}
\usepackage{lmodern,microtype,amsmath,amssymb,bm,graphicx,booktabs,tabularx,array,xcolor}
\usepackage{tikz}
\usetikzlibrary{arrows.meta,positioning,calc}
\usepackage[colorlinks=true,allcolors=blue!55!black]{hyperref}
\newcommand{\UF}{U_{\mathrm F}}
\newcommand{\Fpro}{F_{\mathrm{proc}}}
\newcommand{\Fw}{F_{\mathrm{wc}}}
\newcommand{\Tr}{\operatorname{Tr}}
\newcommand{\Id}{\mathbb I}
\newcommand{\Cay}{\mathcal C}
\makeatletter
\AtBeginDocument{\immediate\write\@auxout{\string\citation{PRXIFormatControl}}}
\makeatother

\begin{document}
\title{Quantum Evolution Networks Learn Long-Time Coherent Dynamics in Driven Spin Registers}
\author{Puran Li}
\affiliation{Department of Physics, City University of Hong Kong, Hong Kong SAR, China}
\affiliation{Shenzhen International Quantum Academy, Shenzhen 518048, China}
\author{Condon Lau}
\affiliation{Department of Physics, City University of Hong Kong, Hong Kong SAR, China}
\author{Xiu-Hao Deng}
\email[Contact author: ]{dengxiuhao@iqasz.cn}
\affiliation{Shenzhen International Quantum Academy, Shenzhen 518048, China}
\affiliation{Shenzhen Branch, Hefei National Laboratory, Shenzhen 518048, China}
\begin{abstract}
Predicting and controlling driven quantum systems requires access to dynamics across physical parameters, initial states, and evolution times. A reusable neural model must capture this dependence while preserving unitary structure and maintaining accuracy over long-time coherent evolution. Here we introduce Quantum Evolution Networks (QENs), which map physical parameters and time to complete propagators for periodically driven three-spin registers. Shared networks learn unitary one-period propagators and compatible within-period paths, enabling predictions through known periodic composition without query-specific numerical integration. For transverse driving, the fifth-percentile fidelity, minimized over input states and sampled times, is 0.9947 over 24 periods on 256 held-out configurations. A leading-order analysis of spectrally resolved endpoint residuals predicts multicycle process error more accurately than quadratic or spectrum-averaged estimates. Interventions at matched single-period infidelity produce orders-of-magnitude differences in multicycle infidelity, demonstrating the importance of residual orientation. Compatible component replacements identify how endpoint and within-period approximations limit reconstruction accuracy. These results establish a reusable neural representation of the studied propagator families and connect its learned residual structure to its observed long-time reliability.
\end{abstract}
\maketitle

\section{Introduction}
Predicting and controlling driven quantum systems requires repeated evaluation of dynamics across physical parameters, initial states, and evolution times. Parameter exploration and control design therefore benefit from models that can answer many related dynamical queries. Machine learning offers a route to encoding this dependence in shared representations\cite{carleo2019rmp}. For tasks requiring coherent evolution of arbitrary input states, the complete propagator provides a common object for these predictions. Learning its parameter dependence can turn a family of dynamical solutions into a reusable model.

Neural representations have been developed for several aspects of quantum dynamics. Neural quantum states encode many-body wavefunctions\cite{carleo2017nqs,glasser2018nqs,gao2017deep,hibatallah2020rnn}, while dynamical extensions describe evolving states and learned updates advance wavepackets\cite{schmitt2020dynamics,vandewalle2025dynamics,gutierrez2022realtime,secor2021propagators}. Propagator learning represents the transformation acting on initial states, with applications to open-system dynamics and time-dependent unitary evolution\cite{zhang2025nqp,guerra2026interpolation}. Function-space approaches, including Fourier neural operators and DeepONet, provide a broader framework for learning solution maps\cite{li2021fno,lu2021deeponet,kovachki2023operator}. Related quantum applications learn effective Floquet generators, observable dynamics, and operator growth\cite{qi2026floquetfno}, as well as propagators across driving protocols\cite{qi2026unp}. These approaches support different forms of reuse, depending on their inputs, learning targets, and imposed physical structure.

For a complete propagator to serve as a reusable dynamical model, physical structure and prediction accuracy must be considered together. In periodically driven systems, approximation errors in a learned one-period propagator can accumulate or partially cancel under repeated composition. Their magnitude alone therefore does not determine multicycle accuracy, while agreement for selected initial states can leave more sensitive states unresolved. This raises the question of how to construct a shared representation of complete propagators and connect its learned errors to the accuracy of its long-time predictions.

Here, we introduce Quantum Evolution Networks (QENs) for prescribed families of periodically driven Hamiltonians. Within each family, a finite parameter vector $\lambda$ specifies the physical configuration, and the overall query map is $(\lambda,t)\mapsto\widehat U_\lambda(t)$. Shared neural weights predict a unitary one-period propagator and a compatible within-period path. Known periodic composition then reconstructs the evolution at queried times, including motion within each period. The resulting operator acts on arbitrary initial states without state-specific fitting. Numerical propagation supplies training labels and evaluation references, while trained-model queries use neural evaluation, unitary decoding, and matrix composition without query-specific integration.

We investigate periodically driven three-spin central-spin registers, comprising an electron coupled to two nuclei through exchange and longitudinal interactions. Longitudinal driving preserves excitation sectors, whereas modulated transverse driving mixes them, providing distinct settings for constructing and testing the representation. The parameter scales are motivated by silicon electron--nuclear registers\cite{morello2020donor,madzik2022,song2026idd}, where long coherence and high-fidelity control have been demonstrated\cite{tyryshkin2012coherence,muhonen2014storing,pla2012singleatom,pla2013nuclear}. The tractable Hilbert space permits complete numerical references and direct minimization of prediction fidelity over all input states. This access allows us to examine both the learned operator and the residual structure underlying its long-time accuracy.

To explain the observed multicycle errors, we resolve actual learned endpoint residuals in the reference Floquet eigenbasis. Building on coherent-error analysis\cite{chen2026interference}, this yields a leading-order estimate of repeated-endpoint process error. Controlled interventions at matched single-period infidelity produce orders-of-magnitude differences in multicycle infidelity, showing that residual orientation carries information about long-time accuracy beyond the single-period error magnitude. Compatible component replacements identify accuracy bottlenecks, while direct worst-input evaluation establishes empirical reliability windows on sampled parameters and times. We further examine training choices and demonstrate reuse of the learned representation in numerically verified control design. The study connects the construction of a reusable complete propagator to a physical explanation of its multicycle accuracy and observed limits.

\section{Quantum Evolution Networks for propagator families}\label{sec:model}
\subsection{A shared map from physical parameters to evolution}
Let $\lambda\in\Lambda$ specify a member of a fixed Hamiltonian family $H(t;\lambda)$. The learning target is the operator-valued solution map
\begin{equation}
\lambda\mapsto U_\lambda(\cdot),\qquad
i\partial_tU_\lambda(t)=H(t;\lambda)U_\lambda(t),\quad U_\lambda(0)=\Id.
\end{equation}
A QEN approximates this map with shared neural weights. A query takes physical parameters and time as inputs and returns a complete propagator, which acts on any pure or mixed initial state without state-specific fitting. The same trained network can be queried across the prescribed Hamiltonian family; the evaluations below assess its accuracy on sampled parameters. Numerical propagation supplies supervised labels and evaluation references. After training, inference uses neural evaluation, prescribed unitary decoding and matrix composition, with no query-specific integration or optimization.

In both physical realizations the interaction contains
\begin{equation}
\begin{aligned}
H_{\rm int}=\sum_{m=1}^2\frac{g_m}{4}
\left[X_mX_e+Y_mY_e+Z_mZ_e\right],\qquad\\
X_mX_e+Y_mY_e=2(\sigma_m^+\sigma_e^-+\sigma_m^-\sigma_e^+).
\end{aligned}
\end{equation}
The exchange and longitudinal terms are tied by the same $g_m$. Longitudinal driving conserves excitation number and permits 19 block coordinates. Transverse driving requires full-space coordinates and uses ten physical and waveform inputs. The Hamiltonians, analytic frames and decoder conventions are specified in Appendix~\ref{app:implementations}. They define two separately trained families at the same three-spin system size.

\subsection{Unitary endpoints, compatible paths and periodic reuse}
For a periodic Hamiltonian $H(t+T;\lambda)=H(t;\lambda)$, write $t=(n+\tau)T$ with integer $n\geq0$ and $0\leq\tau<1$. With the initial drive phase fixed, define
\begin{equation}
\UF(\lambda)=U(T;\lambda),\qquad V(\tau;\lambda)=U(\tau T;\lambda).
\end{equation}
Here $\UF$ is the one-period propagator, referred to as the endpoint, and $V$ is the within-period path. Periodicity gives the exact identity\cite{shirley1965,sambe1973,bukov2015review,goldman2014prx,eckardt2017,oka2019floquet,eckardt2015highfreq}
\begin{equation}
U((n+\tau)T;\lambda)=V(\tau;\lambda)\UF(\lambda)^n.
\label{eq:composition}
\end{equation}
QENs learn both factors and retain this composition rule. Parameter encoders predict Hermitian coordinates, and Cayley decoders produce unitary operators. The path satisfies $\widehat V(0)=\Id$ and $\widehat V(1)=\widehat\UF$ by construction, up to numerical roundoff. The transverse endpoint uses a residual multilayer perceptron (MLP), and the path uses a separate parameter--time attention network. Figure~\ref{fig:method} distinguishes the trained modules from the prescribed periodic composition and analytic frame transformations. These constraints enforce unitarity and the stated endpoint conditions; agreement with the Hamiltonian dynamics is assessed by the evaluations below. The within-period path connects the identity to the endpoint, while a closed Floquet micromotion loop returns to the identity.

The predicted operator supplies $\widehat U(t)|\psi_0\rangle$ for a pure input or $\widehat U(t)\rho_0\widehat U(t)^\dagger$ for a mixed input. Its error can therefore be assessed over all initial states through the relative operator defined in Sec.~\ref{sec:protocol}.

The representation separates two contributions to prediction error. For unitary factors, the spectral norm satisfies
\begin{equation}
\|\widehat V\widehat\UF^{\,n}-V\UF^n\|_2
\leq\|\widehat V-V\|_2+n\|\widehat\UF-\UF\|_2.
\label{eq:error}
\end{equation}
The inequality follows from a telescoping expansion of the endpoint powers. The path enters once, while the endpoint discrepancy enters successive periods. Compatible component replacements with numerical references test which component limits accuracy (Sec.~\ref{sec:components}). The spectral analysis in Sec.~\ref{sec:spectral} further resolves how endpoint errors accumulate or interfere within this worst-case bound.

\begin{figure*}[t]
\centering
\resizebox{0.9\textwidth}{!}{%
\begin{tikzpicture}[font=\footnotesize,>=Latex,
net/.style={draw=blue!65!black,fill=blue!4,rounded corners,align=center,minimum height=10mm},
fixed/.style={draw=black!60,fill=black!3,rounded corners,align=center,minimum height=10mm},
arr/.style={->,thick}]
\node[anchor=west,font=\bfseries] at (-7.2,1) {(a) Learning a dynamics family};
\node[fixed,text width=29mm] (input) at (-5.5,0) {Physical inputs\\$\lambda\in\Lambda$};
\node[net,text width=40mm] (shared) at (-1.1,0) {Shared trained weights\\$\lambda\mapsto[\widehat U_F,\widehat V(\cdot)]$};
\node[fixed,text width=49mm] (full) at (4.4,0) {Known periodic composition\\$\widehat U((n+\tau)T)=\widehat V(\tau)\widehat U_F^n$};
\draw[arr] (input)--(shared);\draw[arr] (shared)--(full);
\draw[black!25] (-7.2,-.9)--(7.2,-.9);
\node[anchor=west,font=\bfseries] at (-7.2,-1.4) {(b) Z: symmetry-adapted representation};
\node[fixed,text width=25mm] (zin) at (-5.7,-3) {Six physical\\parameters};
\node[net,text width=36mm] (zep) at (-1.7,-2.3) {Residual MLP\\19 endpoint coordinates};
\node[net,text width=36mm] (zpath) at (-1.7,-4) {Chebyshev features\\Attention encoder\\Harmonic coefficients};
\node[fixed,text width=39mm] (zuf) at (3,-2.3) {Block-Cayley decoder\\$\widehat U_F$};
\node[fixed,text width=52mm] (zv) at (3.8,-4.2) {Harmonics at $\tau$; block Cayley\\Residual loop $\widehat R(\tau)$\\$\widehat V_Z=U_{\rm fast}\widehat R\widehat U_F^\tau$};
\draw[arr] (zin.east)--++(.25,0)|-(zep.west);\draw[arr] (zin.east)--++(.25,0)|-(zpath.west);
\draw[arr] (zep)--(zuf);\draw[arr] (zpath)--(zv);\draw[arr] (zuf.south)--(zuf.south|-zv.north);
\draw[black!25] (-7.2,-5.3)--(7.2,-5.3);
\node[anchor=west,font=\bfseries] at (-7.2,-5.85) {(c) X: full-space representation};
\node[fixed,text width=29mm] (xin) at (-5.5,-7.2) {Ten parameters\\50 scaled features};
\node[net,text width=43mm] (xep) at (-1,-7.2) {Residual MLP + feature skip\\$3\times63$ bounded coordinates};
\node[fixed,text width=43mm] (xuf) at (4,-7.2) {$U_{\rm ref}$ + three Cayley layers\\Unitary endpoint $\widehat U_F$};
\draw[arr] (xin)--(xep);\draw[arr] (xep)--(xuf);
\node[fixed,text width=29mm] (time) at (-5.5,-9.7) {Parameter\\features (50)\\Time features (9)};
\node[net,text width=43mm] (xp) at (-1,-9.7) {Two projected tokens\\Attention; pool + features\\$d_\phi=\tau(1-\tau)f_\phi\in\mathbb R^{64}$};
\node[fixed,text width=43mm] (xf) at (4,-9.7) {Full Pauli / Cayley decoder\\$c(\tau)=\tau c_F+d_\phi$\\Reference + frame $\to\widehat V_X$};
\draw[arr] (time)--(xp);\draw[arr] (xp)--(xf);
\draw[arr] (xuf.south)--node[right,font=\scriptsize,align=left]{inverse Cayley\\$c_F$ at inference}(xf.north);
\node[align=left,font=\scriptsize,anchor=west] at (-7.1,-11) {Training: reference $c_F$ defines path labels. Inference: predicted $c_F$ enters the decoder.\\The path encoder receives parameters and time, not $c_F$. Blue: trained modules. Gray: prescribed operations.};
\end{tikzpicture}}
\caption{Structure-constrained learning of a propagator family. (a) Shared weights map physical parameters to a one-period propagator and a within-period path; periodic composition produces the queried operator. (b) Longitudinal driving uses block coordinates and a harmonic residual loop. (c) Transverse driving uses a residual MLP and three full-space Cayley factors for the endpoint, together with a parameter--time path network. Blue modules are trained; gray modules are prescribed transformations. Reference endpoint coordinates define path supervision, and predicted endpoint coordinates enter the path decoder at inference. This interface supports the component replacements in Sec.~\ref{sec:components}.}
\label{fig:method}
\end{figure*}
\section{Numerical scope and empirical reliability criteria}\label{sec:protocol}
\subsection{Parameter domains and numerical references}
The longitudinal inputs are $\lambda_Z=(\omega_1,\omega_2,g_1,g_2,\omega_d,A)$, with each coordinate varying by $\pm15\%$ around $\lambda_{Z,0}=(74.82,24,20,10,3000,32160)$. These coefficients are angular rates in $\mu\mathrm{s}^{-1}$, giving $T_Z=2\pi/\omega_d$. One thousand nominal periods span $2.094\,\mu\mathrm{s}$. The driving frequency is high relative to the nuclear and coupling scales, while the analytical rotation retains the full modulation amplitude.

The transverse inputs are $(g_1,g_2,\Delta_e,\Omega_0,a_1,b_1,a_2,b_2,\beta_1,\beta_2)$. Frequency inputs are stored in MHz and converted to angular rates in the Hamiltonian. The fixed ordinary frequencies are $f_1=74.82$ MHz, $f_2=24$ MHz, $f_c=32160$ MHz, and $f_\nu=24$ MHz. Thus $T_X=1/24\,\mu\mathrm{s}$. The sampled intervals are $g_1,g_2\in[10,40]$ MHz, $\Delta_e\in[-2,2]$ MHz, and $\Omega_0\in[8,14]$ MHz. Scrambled Sobol sampling is combined with uniform sampling in the amplitude-coefficient ball $\|\mathbf a\|_2\leq1/(2\sqrt2)$ and phase-coefficient disk $\|\bm\beta\|_2\leq0.25$. These joint constraints give $A(t)\geq\Omega_0/2$.

The longitudinal endpoint uses 65,536 training parameters and its path uses 32,768; both have 4,096 validation and 8,192 test members. A 2,048-member held-out set is evaluated over the full time grid. Each transverse component uses a 16,384-member training pool and a 2,048-member validation pool. Training and checkpoint selection are specified in Appendix~\ref{app:training}. For the transverse family, evaluation over the full time grid uses 2,048 held-out configurations, with a fixed 256-member subset for direct spectral evaluation of worst-input fidelity. The primary transverse endpoint is selected by repeated-endpoint validation fidelity at 100 and 24 periods; its training loss is single-period. Both neural components remain frozen during evaluation. The comparison, continuation and timing experiments use the explicitly identified checkpoints in Table~\ref{tab:modelmap}.

Here, held-out denotes parameters excluded from gradient training and checkpoint selection. The transverse test parameters have been reused across successive model and diagnostic studies; each reported experiment uses the fixed models and validation-only selection rules specified for that experiment. The studies over the full time grid and the spectral studies retain the same member identities, enabling paired comparisons. Table~\ref{tab:modelmap} records the model and evaluation scope for each result.

Numerical integration supplies training labels and evaluation references. Transverse evaluation uses a fourth-order commutator-free integrator\cite{blanes2006cf} with 4,096 and 8,192 steps per period, Richardson extrapolation, and polar projection. The full process evaluation exports 65 within-period nodes, including both boundaries. The spectral subset adds 16 intermediate phases. Finer-resolution checks and the endpoint--path interface tests are specified in Appendix~\ref{app:reference}. We refer to numerical propagation as the ODE reference throughout the figures.

\subsection{Process and worst-input metrics}
\begin{sloppypar}
Process and worst-input errors describe complementary properties of an approximate quantum operation\cite{gilchrist2005distance,sanders2016bounds}. These operator-level measures complement experimental characterization protocols---\allowbreak randomized bench\-marking\cite{emerson2005rb,knill2008rb,magesan2011rb,wallman2014rb}, gate-set tomography\cite{merkel2013gst,blumekohout2017gst}, and direct fidelity estimation\cite{flammia2011dfe}---which bound gate errors from observed data. For $d=8$, the phase-invariant process fidelity is
\end{sloppypar}
\begin{equation}
\Fpro(U,\widehat U)=\frac{|\Tr(U^\dagger\widehat U)|^2}{d^2}.
\end{equation}
To test a fidelity requirement for every input state, we evaluate the least accurately evolved input directly through
\begin{equation}
\Fw(U,\widehat U)=\min_{\|\psi\|=1}|\langle\psi|U^\dagger\widehat U|\psi\rangle|^2.
\label{eq:worstinput}
\end{equation}
For a relative unitary $W=U^\dagger\widehat U$, this equals the squared distance from the origin to the convex hull of its eigenvalues. If their shortest containing circular arc has width $a<\pi$, then $\Fw=\cos^2(a/2)$; otherwise $\Fw=0$. The spectral calculation minimizes over all input states for the supplied operators. Parameter configurations and times are sampled as specified below.

Transverse worst-input fidelity is evaluated directly from the relative spectrum. For the longitudinal process data, we use the conservative lower bound derived in Appendix~\ref{app:bound}. Appendix~\ref{app:channel} gives the corresponding channel-distance measures and distributions.

\subsection{Time windows and parameter coverage}
For an integer horizon $H$, the transverse grid $\mathcal G_H$ contains $(n+\tau)T_X$ for $0\leq n<H$ and within-period nodes $0\leq\tau<1$, together with $HT_X$. For either fidelity metric, define
\begin{equation}
S_F(\lambda,H)=\min_{t\in\mathcal G_H}F(U(t;\lambda),\widehat U(t;\lambda)).
\end{equation}
We write $S_{\mathrm{proc}}$ and $S_{\mathrm{wc}}$ for the process and worst-input choices of $F$, and summarize these per-parameter scores by their median, fifth percentile (p05), and minimum. At a chosen fidelity threshold, empirical parameter coverage is the fraction of evaluated configurations whose scores meet or exceed that threshold. The empirical reliable horizon $H^*$ is the last evaluated integer $H$ meeting the specified coverage. At a chosen tolerance, it reports how long the required fraction of sampled configurations passes at every queried time. For $F=\Fw$, each passing configuration meets the threshold for every input state on that grid. The 81-node spectral grid spans 80 distinct phases per period with endpoints identified, giving 8,001 query times through $100T_X$. Bootstrap intervals use 10,000 resamples of parameter members with a fixed seed; they describe parameter-sample variation conditional on the frozen models. The reported horizon is an empirical property of the evaluated parameters and grid. For integer-endpoint comparisons, $F_h$ denotes the process fidelity of the repeated endpoint at $hT$.
\section{Reliability of complete evolution across the transverse family}\label{sec:mainresults}
\subsection{Worst-input fidelity and sampled coverage}
We first assess the time window over which the frozen transverse QEN remains accurate across its ten physical inputs. On 2,048 held-out parameter configurations, the fifth-percentile worst-time process fidelity was 0.9976 over $24T_X=1\,\mu\mathrm{s}$, with median 0.9993. The sampled time grid included each period's internal dynamics and its endpoint. Extending the window to $100T_X=4.17\,\mu\mathrm{s}$ gave p05 0.9777 and median 0.9948 (Table~\ref{tab:xprocess}). Every prediction used the same endpoint and path networks.

Direct spectral evaluation tests whether this accuracy holds for every input state. On the 256-member subset, worst-input fidelity over $24T_X$ had p05 0.9947, with a parameter-bootstrap 95\% interval of $[0.9932,0.9957]$. Of these members, 252 (98.4\%) retained worst-input fidelity of at least 0.99 throughout the sampled window. The minimum score was 0.9732. Thus the threshold holds for all inputs at all queried times on those 252 configurations, with the parameter quantiles describing variation across the tested family.

\begin{figure*}[t]\centering
\includegraphics[width=0.90\textwidth]{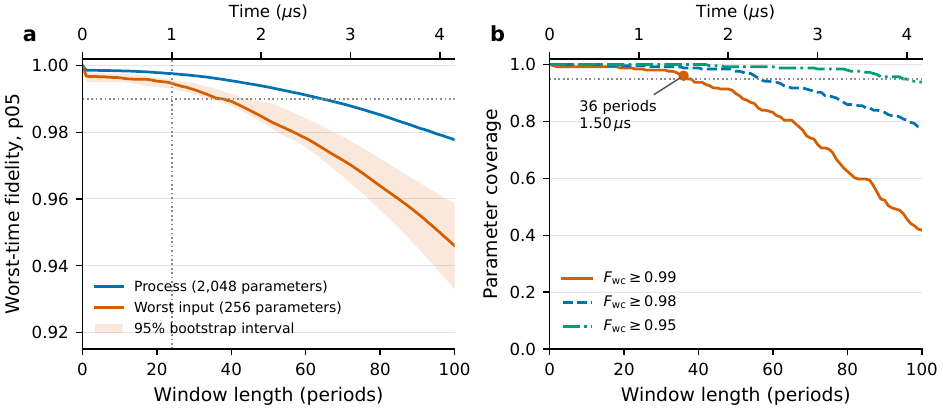}
\caption{Empirical reliability windows under transverse driving. (a) Fifth-percentile worst-time process fidelity on 2,048 held-out parameters and worst-input fidelity on a 256-member subset. The shadow is a pointwise 95\% parameter-bootstrap interval for the worst-input p05. Dotted lines mark $24T_X$ and fidelity 0.99. (b) Fractions of the spectral subset exceeding three worst-input fidelity thresholds throughout each evaluated window. At threshold 0.99 and coverage 95\%, the last passing window is $36T_X=1.50\,\mu\mathrm{s}$. These windows use the parameter samples and time grids specified in Sec.~\ref{sec:protocol}.}
\label{fig:worst}
\end{figure*}

\begin{table*}[t]\centering\small
\caption{Accuracy over the full time grid of the transverse QEN. Process scores use 2,048 held-out parameters; worst-input scores use the 256-member spectral subset. Each score is minimized over the corresponding time grid before taking parameter quantiles.}\label{tab:xprocess}
\begin{tabular}{rrrrrrr}\toprule
& & \multicolumn{3}{c}{Process fidelity} & \multicolumn{2}{c}{Worst-input fidelity}\\
\cmidrule(lr){3-5}\cmidrule(lr){6-7}
$H$ & Time ($\mu$s) & Median & p05 & Minimum & Median & p05\\\midrule
1 & 0.0417 & 0.9997 & 0.9986 & 0.9943 & 0.9993 & 0.9969\\
24 & 1.00 & 0.9993 & 0.9976 & 0.9878 & 0.9986 & 0.9947\\
100 & 4.17 & 0.9948 & 0.9777 & 0.8112 & 0.9880 & 0.9460\\\bottomrule
\end{tabular}
\end{table*}

Figure~\ref{fig:worst} converts the accuracy curves into a criterion for repeated use. At a worst-input fidelity threshold of 0.99, at least 95\% empirical coverage holds through 36 periods ($1.5\,\mu\mathrm{s}$, 246/256 members) and breaks at 37 periods, where coverage falls to 94.9\%. The resulting horizon states when this tolerance ceases to hold for the required fraction of the evaluated configurations. The distributions in Appendix~\ref{app:channel} further resolve process error, worst-input error, and channel distinguishability on the same subset.

\subsection{Boundary sensitivity and reference resolution}
In a separate coupling-boundary test at $24T_X$, all eight interior anchors, 27 of 32 face members and 11 of 32 corner members retained worst-input fidelity at least 0.99. Their p05 scores were 0.9918, 0.9831 and 0.9343, respectively. The face and corner members share their other physical inputs with paired interior anchors. The poorer boundary scores show why these locations require separate assessment when defining the range of reliable use. Appendix~\ref{app:newreference} gives the paired construction and decoder-conditioning checks.

Reference refinement on 32 preselected random, tail, boundary and control members changed worst-input fidelity by at most $3.0\times10^{-6}$ at 24 periods and $4.2\times10^{-5}$ at 100 periods, without changing threshold decisions on those members. This check supports the reported numerical resolution within that subset.

Appendix~\ref{app:trajectories} illustrates the use of the same learned operators to predict populations and coherences and shows how error varies across parameter samples. The worst-input scores above establish the fidelity criterion over all initial states. The next section examines how the measured endpoint residual produces its multicycle error.
\section{Spectral structure of learned endpoint errors}\label{sec:spectral}
\subsection{Resolving the learned error in the Floquet eigenbasis}
The reliability window records the outcome of repeated prediction. To explain the endpoint contribution, we examine how its approximation error is distributed in operator space at each parameter configuration. The numerical endpoint $U$ and learned endpoint $\widehat U$ define a residual whose magnitude and direction determine its repeated action. We apply the established conjugation-sum analysis of coherent accumulation and error interference\cite{chen2026interference} to these actual neural residuals. Let
\begin{equation}
 U^\dagger\widehat U=e^{-iE},\qquad U=Q\,\operatorname{diag}(e^{-i\theta_a})Q^\dagger,
 \label{eq:spectralerror}
\end{equation}
where $E$ is the principal Hermitian error generator. In the phase-invariant calculations below, we remove its identity component and retain $E$ for the resulting traceless generator; the relative-unitary identity is then understood up to an overall phase. To first order in the endpoint error, the generator after $n$ repetitions is a sum of conjugated copies of $E$. In the Floquet eigenbasis, each matrix element is weighted by
\begin{equation}
\begin{aligned}
 w_n(\Delta_{ab})=\left|\sum_{k=0}^{n-1}e^{ik\Delta_{ab}}\right|^2,\\
 \quad\Delta_{ab}=\theta_a-\theta_b.
\end{aligned}
\end{equation}
The resulting error-resolved prediction of leading-order process infidelity is
\begin{equation}
 P_2(n)=\frac{1}{d}\sum_{a,b}w_n(\Delta_{ab})\,
 |(Q^\dagger E Q)_{ab}|^2,\qquad d=8.
 \label{eq:spectralprediction}
\end{equation}
For components commuting with $U$, the weight is $n^2$. Components joining distinct eigenphases acquire oscillatory weights. At a fixed nonzero circular gap, these weights are bounded in $n$, although a near-degenerate gap gives approximately quadratic growth while its accumulated phase remains small. Equation~\eqref{eq:spectralprediction} therefore identifies which residual contributions accumulate strongly at a given horizon by combining their magnitudes with the reference eigenphase gaps.

\begin{figure*}[t]\centering
\includegraphics[width=0.90\textwidth]{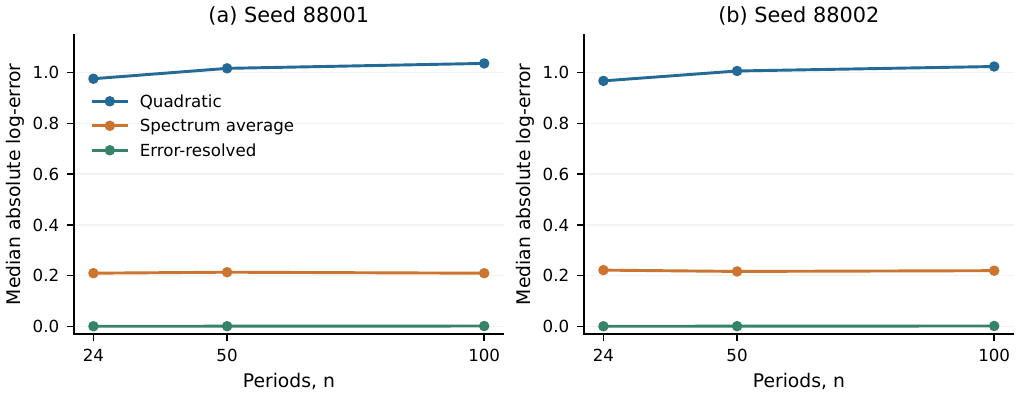}
\caption{Predicting repeated-endpoint error from the learned single-period residual. Each panel uses the original endpoint for one training seed on 256 common-test members. The vertical axis is the median absolute difference between predicted and measured log$_{10}$ process infidelity, evaluated at the indicated period count. The quadratic estimate, spectrum-averaged estimate, and error-resolved prediction are $P_0$, $P_1$, and $P_2$, respectively. Exact matrix powers supply the measured errors.}\label{fig:spectralprediction}
\end{figure*}

We compared this error-resolved prediction with the quadratic estimate $P_0(n)=n^2[1-\Fpro(U,\widehat U)]$ and a spectrum-averaged estimate $P_1$, which averages over residual orientation at fixed reference spectrum (Appendix~\ref{app:spectral}). All three estimates were tested against exact matrix powers on the fixed 256-member subset. Figure~\ref{fig:spectralprediction} shows the median absolute logarithmic prediction error for the two independently trained endpoints. Resolving the learned error in the Floquet basis closely follows the measured process infidelity at 24, 50, and 100 periods. At 100 periods, the median absolute log$_{10}$ errors of $P_2$ are 0.0015 and 0.0017; the corresponding rank correlations exceed 0.9998. The spectrum-averaged estimate captures part of the reduction from quadratic accumulation, while the error-resolved prediction captures differences among parameter members. This improvement shows the information gained by retaining the actual residual orientation. Appendix~\ref{app:spectral} specifies the small-error expansion and its evaluation regime. The spectral analysis uses reference operators to predict repeated-endpoint process error; the worst-time worst-input criterion is evaluated separately in Sec.~\ref{sec:mainresults}.

\subsection{Interventions at matched single-period accuracy}
We isolated the role of residual orientation through a controlled intervention. At each parameter member, the original error generator was separated into its component commuting with $U$ and its complementary mixing component. Within exactly degenerate eigenspaces the full block was retained. Each component was independently rescaled so that the perturbed endpoint matched the original single-period process infidelity. Both perturbed endpoints were propagated by direct matrix powers using the same reference $U$. These constructed operators test the sensitivity to residual orientation within measured residuals of frozen networks.

\begin{figure*}[t]\centering
\includegraphics[width=0.90\textwidth]{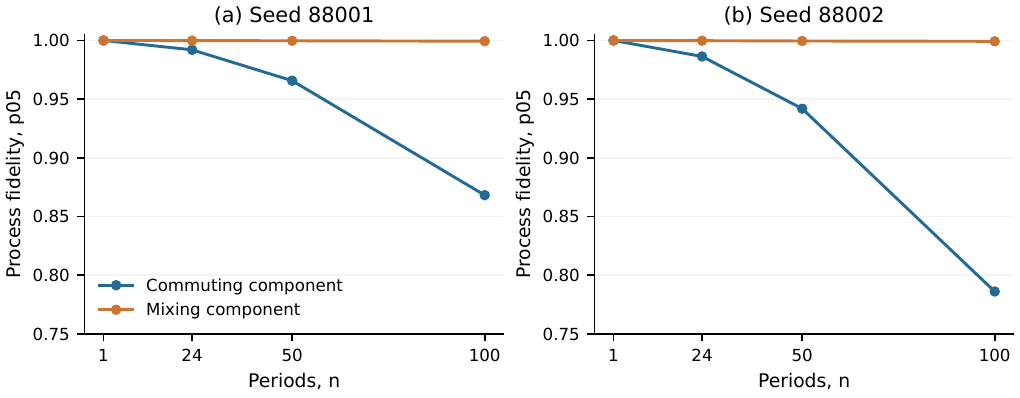}
\caption{Residual-orientation intervention at matched single-period process accuracy. Original learned residuals are projected into commuting and mixing components relative to the reference endpoint, then separately rescaled to match the same member's single-period infidelity. Curves show process-fidelity p05 on 254 paired members for seed 88001 and 255 for seed 88002, drawn from the same fixed 256-member subset. Both directions within a panel use identical member indices. Matching, degeneracy, and admissibility rules are fixed before evaluation and given in Appendix~\ref{app:spectral}.}\label{fig:spectraldirection}
\end{figure*}

Both directions use exactly the same paired members: 254 of 256 for the first endpoint and 255 for the second, after the joint matching and scale checks. The two directions exhibited sharply different multicycle behavior despite their matched single-period scores (Fig.~\ref{fig:spectraldirection}). For these paired interventions, the commuting component accumulated substantially larger errors, while the mixing component retained high process fidelity through 100 periods. At 100 periods, the corresponding process-infidelity p95 values differed by more than 200-fold in both subsets. Because single-period infidelity is matched member by member, this comparison isolates the effect of residual orientation in the Floquet eigenbasis. It establishes why an accuracy requirement for repeated use must be checked over the intended horizon. Whether the diagnostic also improves training is tested next by continuing the neural models under different objectives.

\section{Representation choices and training outcomes}\label{sec:representation}
We next apply the multicycle criteria to assess changes in training and representation. The comparisons test whether a different objective improves the reliability window, whether encoder choice changes the accuracy--cost trade-off, and which component limits the complete prediction.

\subsection{Spectral weighting and ordinary continuation}
To test whether the spectral analysis improves optimization, we continued each of the two trained endpoints for 2,000 updates under three objectives: A retained the single-period coordinate objective, B incorporated normalized spectral weighting, and C incorporated directly evaluated multicycle process error. The branches shared initial weights, parameter batches, optimizer settings and update counts. Equal update counts fix the number of training steps; the objectives can have different per-step costs. Validation evaluated checkpoints at updates 0, 250, \ldots, 2,000, selecting the integer-period reliable horizon first and using long-time tail fidelity to break ties. The selected endpoints were evaluated on 2,048 common-test parameters and then combined with the frozen path network for the 256-member audit over the full time grid.

\begin{figure*}[t]\centering
\includegraphics[width=0.90\textwidth]{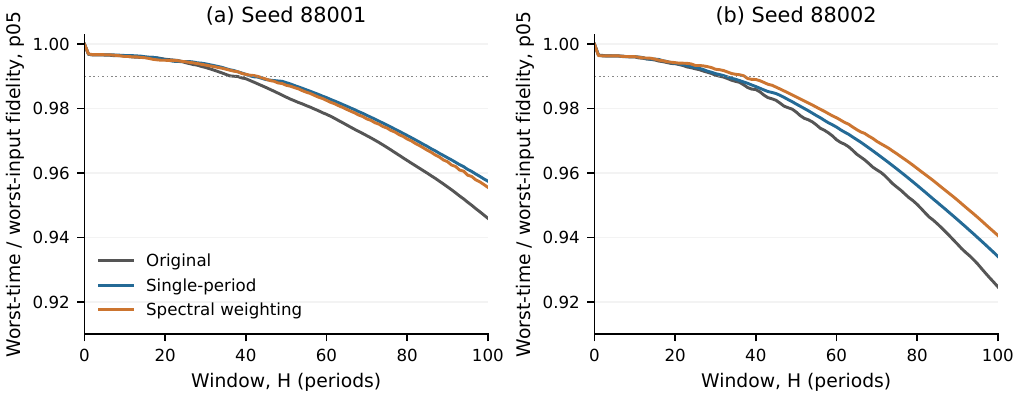}
\caption{Worst-time worst-input accuracy of the original and continued QENs on 256 common-test members. Each curve uses a validation-selected endpoint and the same frozen path network, with the endpoint prediction inserted into that path's boundary condition. Fidelity is minimized across all queried times and input states within each window. C coincides with the original curve because validation selects update 0.}\label{fig:spectraltraining}
\end{figure*}

Continuation increased the reliable horizon from 36 to 41 periods for the first endpoint under both A and B. For the second endpoint it increased from 30 to 32 periods under A and to 36 under B (Table~\ref{tab:spectraltraining}). At 100 periods, A gave the higher p05 for the first endpoint and B for the second. Validation selected the original checkpoint for C in both runs. Both A and B extend the original model's reliability window, with their relative performance varying across runs and metrics. Figure~\ref{fig:spectraltraining} shows the curves over the full time grid, including the original model.

At 100 periods, paired parameter-bootstrap 95\% intervals for the B-minus-A differences in worst-time worst-input fidelity p05 included zero for both seeds (Appendix~\ref{app:spectral}). For seed 88001, the difference was $-0.0019$, with interval $[-0.0102,0.0070]$. Its endpoint-only process-fidelity p05 also favored A: 0.9828 versus 0.9807 for B (Table~\ref{tab:continuedendpoint}). The measured reliability windows provide a direct way to assess each selected continuation, but the two runs do not establish a consistent advantage for spectral weighting. Selection of the original checkpoint for C describes this continuation protocol's outcome. The quantitative success of the spectral analysis and the effectiveness of a training objective remain separate results.

\begin{table*}[t]\centering\small
\caption{Evaluation over the full time grid after validation-selected continuation. Scores $S_{\mathrm{wc}}(\lambda,H)$ are minimized over the 81-node-per-period multicycle grid and over all input states before taking parameter p05 on the 256-member subset. $H^*$ uses fidelity 0.99 and at least 95\% parameter coverage. C selects update 0 and has the same scores as O.}\label{tab:spectraltraining}
\begin{tabular}{llrrrr}\toprule
Seed & Objective & Update & $H^*$ & $S_{\mathrm{wc}}(\lambda,24)$ p05 & $S_{\mathrm{wc}}(\lambda,100)$ p05\\\midrule
88001 & O: original & 0 & 36 & 0.9947 & 0.9460\\
 & A: single-period & 250 & 41 & 0.9946 & 0.9575\\
 & B: spectral & 250 & 41 & 0.9947 & 0.9556\\
 & C: multicycle & 0 & 36 & 0.9947 & 0.9460\\\midrule
88002 & O: original & 0 & 30 & 0.9927 & 0.9247\\
 & A: single-period & 500 & 32 & 0.9930 & 0.9341\\
 & B: spectral & 250 & 36 & 0.9939 & 0.9406\\
 & C: multicycle & 0 & 30 & 0.9927 & 0.9247\\\bottomrule
\end{tabular}
\end{table*}

\subsection{Encoder choice under a shared unitary decoder}\label{sec:xablation}
The residual MLP and attention encoders share the same three-factor full-space Cayley decoder, which allows accuracy and update time to be compared under a common schedule. Two independent MLP runs achieved held-out repeated-endpoint $F_{24}$ p05 of 0.9987 and 0.9984, and $F_{100}$ p05 of 0.9795 and 0.9740. These integer-period metrics evaluate the endpoint map directly. The endpoint used for evaluation over the full time grid was selected by its validation score before being combined with the path network.

With the same decoder, parameter-matched attention encoders gave $F_{24}$ p05 of 0.9985 and 0.9988. The MLP used approximately one-third of their cumulative update time under the shared training schedule. The accuracy ordering reversed between seeds, whereas the training-time ordering was consistent. Under this schedule, the MLP offers lower update cost with comparable accuracy across the two runs. Appendix~\ref{app:xablation} reports the full comparison and the four-seed study that motivated the full-space decoder.

\subsection{Component replacement within the shared endpoint--path representation}\label{sec:components}
The endpoint--path interface permits compatible component replacements that preserve the initial condition, endpoint consistency and periodic composition. We replace the learned endpoint, residual loop, or path residual by its numerical reference and measure the resulting reconstruction error. Endpoint replacements also update the path connection to maintain endpoint consistency. These comparisons identify accuracy bottlenecks within the chosen reconstruction, whose errors need not add independently.

\paragraph{Longitudinal reconstruction.}
The longitudinal realization applies the endpoint--path representation in conserved excitation sectors. Across 2,048 held-out members, its worst-time process fidelity had p05 0.9990 and median 0.9999. The grid covers 65 within-period nodes at period indices $n=0,\ldots,1000$, ending at $1001T_Z$. The nominal physical duration is $2.10\,\mu\mathrm{s}$, with the large cycle count arising from the rapid longitudinal drive.

\begin{figure*}[t]\centering
\includegraphics[width=0.90\textwidth]{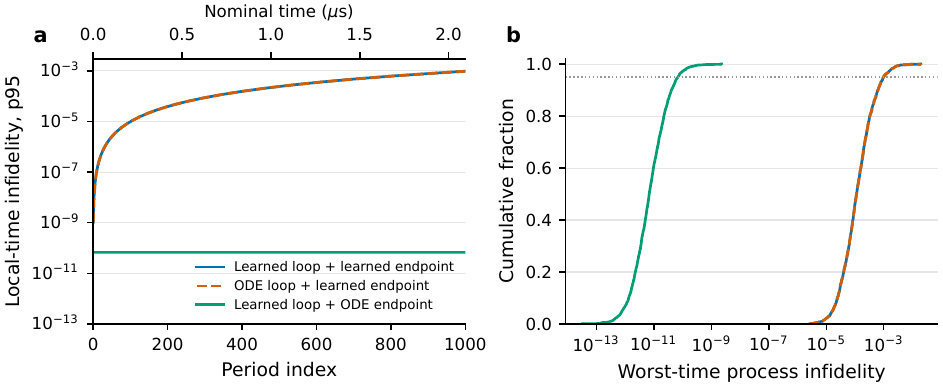}
\caption{Complete longitudinal dynamics and component replacement on 2,048 held-out members. (a) The 95th-percentile process infidelity after minimizing fidelity over local time within each period. The upper axis uses the nominal drive frequency. (b) Distributions of each member's maximum infidelity over the full time grid. Replacing the learned loop by the ODE loop leaves the endpoint-dominated error nearly unchanged; replacing the endpoint resolves the much smaller loop contribution relative to the fixed ODE reference.}
\label{fig:z}
\end{figure*}

Component replacement separated the two learned contributions. Pairing the ODE loop with the learned endpoint reproduced the full model's long-time statistics, leaving the endpoint-dominated error essentially unchanged. The converse pairing---the learned loop with the ODE endpoint---isolates the loop's own contribution: relative to the fixed reference, its 95th-percentile worst-time process infidelity was $6.8\times10^{-11}$, with maximum $2.2\times10^{-9}$. The endpoint consequently controlled long-time error in this realization, consistent with its repeated use in Eq.~\eqref{eq:error}.

The longitudinal study also compared output coordinates and endpoint approximations. At 16,384 training members, the 19-coordinate block model gave a two-seed mean repeated-endpoint $F_{1000}$ p05 of 0.9949, compared with 0.9932 for a parameter-matched 63-coordinate representation. On a common 256-member subset evaluated over the full time grid, a Chebyshev endpoint combined with the learned loop achieved process-infidelity p95 of $1.3\times10^{-6}$, compared with $4.1\times10^{-4}$ for the neural endpoint. Appendix~\ref{app:zcomparisons} gives the distributions and coordinate-loss definitions. The comparison identifies an effective component replacement for this family: structured regression improves accuracy while retaining the learned loop and updating its connection to the replacement endpoint.

\paragraph{Transverse reconstruction.}\label{sec:xhybrid}
We apply the same component replacement analysis to the transverse family on the 256 common-test members and 81 within-period nodes used for the audit over the full time grid. Endpoint replacement updates both the path's boundary coordinates and the repeated-period operator, preserving the initial and endpoint conditions. Numerical path residuals are extracted from the ODE reference in the same Cayley coordinates as the learned path (Appendix~\ref{app:implementations}). All neural weights and the validation-selected regression endpoint remain frozen, so the comparisons test the effect of compatible component replacements in a fixed reconstruction.

\begin{figure*}[t]\centering
\includegraphics[width=0.90\textwidth]{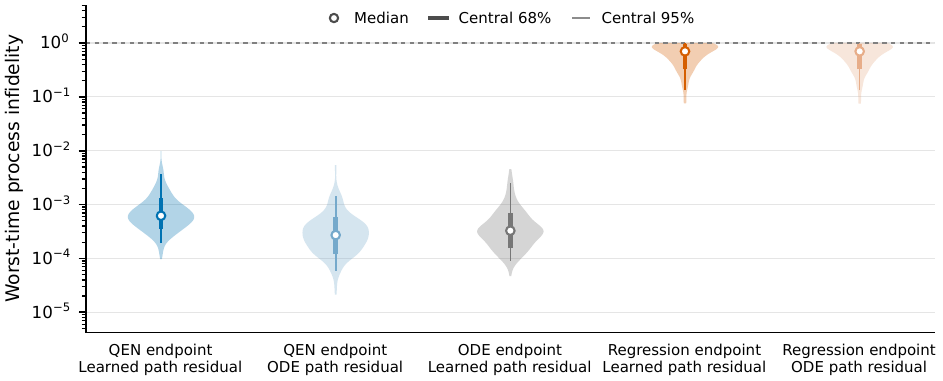}
\caption{Transverse component replacement on 256 common-test members over $24T_X=1\,\mu\mathrm{s}$. Each member contributes its maximum process infidelity over the augmented time grid. Endpoint replacements update both boundary coordinates and the repeated endpoint. Learned and ODE path residuals denote the within-period coordinates $\mathbf d$ in Eq.~\eqref{eq:xpath}. Shadows show the distribution of test-set data points; the regression distributions end near the physical infidelity bound of one, marked by the dashed line. Open circles mark medians, thick lines span the 16th--84th percentiles, and thin lines span the 2.5th--97.5th percentiles. All errors use the same fixed ODE reference.}\label{fig:xhybrid}
\end{figure*}

Over $24T_X=1\,\mu\mathrm{s}$, the full QEN gave worst-time process-infidelity p95 of $2.37\times10^{-3}$ (Fig.~\ref{fig:xhybrid}). Replacing its path residual with the numerical residual reduced this to $1.28\times10^{-3}$; replacing its endpoint with the numerical endpoint gave $1.51\times10^{-3}$. Both learned components thus contribute to the error over this window. At 100 periods, the full QEN and numerical-path reconstruction gave p95 values of $2.19\times10^{-2}$ and $2.04\times10^{-2}$, while the numerical-endpoint reconstruction remained at $1.51\times10^{-3}$. The endpoint approximation increasingly controls the long-time growth, connecting the component comparison to the spectral analysis in Sec.~\ref{sec:spectral}.

Figure~\ref{fig:xhybrid} also includes a validation-selected polynomial endpoint paired with either path residual; its approximation and results are examined in Sec.~\ref{sec:qicomparison}. The component tests imply different priorities for improving the two realizations. Refining the endpoint addresses the dominant longitudinal error. In the transverse case, both approximations matter over the target window, while improving the endpoint becomes increasingly important at longer horizons. This dependence on the family and horizon is resolved by the component replacements evaluated over the full time grid.

\subsection{Alternative transverse endpoint representations}\label{sec:qicomparison}
The growing endpoint contribution motivates examining alternative transverse endpoint approximations under the same repeated-use metrics. We compared QEN with an adapted implementation of the Floquet-Hamiltonian learning approach of Qi et al.\cite{qi2026floquetfno}. The comparator maps time-dependent Hamiltonian coefficients to effective Floquet-Hamiltonian coefficients and constructs $\widehat U_{\mathrm F}=\exp[-iT_X\widehat H_{\mathrm F}(0)]$. Both models were evaluated against the same reference endpoints on 2,048 held-out configurations, with the physical starting phase fixed at zero. The comparison records two trained implementations for the same parameter-to-endpoint task, with distinct supervision and optimization schedules specified below.

The single-period process-infidelity p95 was $1.7\times10^{-5}$ for QEN and $1.5\times10^{-1}$ for the tested Floquet-Hamiltonian FNO (Table~\ref{tab:qicomparison}). The comparator's median single-period fidelity was 0.9970, while its p05 was 0.8503, resolving a substantial difference between typical and lower-tail accuracy. QEN achieved a median single-period infidelity of $4.7\times10^{-6}$. The difference was already present in the learned endpoint, before any repeated-period reconstruction.

Applying the same periodic reuse to both predictions exposed the consequence for long-time coherence. At 24 periods, process-fidelity p05 was 0.9988 for QEN and 0.0310 for the comparator; at 100 periods, the corresponding values were 0.9793 and 0.0071. These scores show why typical single-period accuracy should be accompanied by lower-tail and multicycle evaluation. The FNO received 14,000 updates and QEN 32,000. Validation selection used effective-generator coefficient MSE for FNO and repeated-endpoint fidelity for QEN. Validation selected update 13,000 for FNO and 31,000 for QEN; these checkpoint indices differ from the total update counts. Appendix~\ref{app:qicomparison} specifies the adapted implementations. The measured gap concerns these fitted maps; attributing it to architecture or interaction structure would require a controlled training comparison. Table~\ref{tab:xprocess} separately evaluates the QEN's reconstruction over the full time grid.

\begin{table*}[t]\centering\small
\caption{Endpoint comparison on 2,048 held-out transverse configurations. Single-period errors are process-infidelity quantiles; later-time entries are process-fidelity fifth percentiles at the specified integer period. Both methods use the fixed starting phase $t=0$. QEN here denotes the endpoint run in the comparison experiment; results over the full time grid for the manuscript model are reported separately in Table~\ref{tab:xprocess}.}\label{tab:qicomparison}
\begin{tabular}{lrr}\toprule
Metric & QEN & Floquet-Hamiltonian FNO\\\midrule
Single-period infidelity, median & $4.7\times10^{-6}$ & $3.0\times10^{-3}$\\
Single-period infidelity, p95 & $1.7\times10^{-5}$ & $1.5\times10^{-1}$\\
$F_{24}$, p05 & 0.9988 & 0.0310\\
$F_{100}$, p05 & 0.9793 & 0.0071\\\bottomrule
\end{tabular}
\end{table*}

A complementary polynomial endpoint regressed the transverse endpoint's 63-dimensional reference-residual Cayley coordinates. Validation selected degree three with 286 features and ridge regularization $0.01$. On 2,048 test parameters its repeated-endpoint process-fidelity p05 was 0.9610 at one period, 0.0020 at 24 periods, and $8.1\times10^{-4}$ at 100 periods. Exact-coordinate decoding reconstructed the reference to numerical precision, locating the approximation error in the fitted parameter dependence. In the component comparison over the full time grid, this polynomial endpoint gave median worst-time infidelity 0.699 with either the learned or numerical path residual over 24 periods (Fig.~\ref{fig:xhybrid}). The close distributions identify the fitted endpoint as the accuracy bottleneck in these reconstructions. Its performance characterizes the tested regression family; the longitudinal comparison shows that a structured endpoint can be effective in a different parameter regime.

\section{Reuse for online queries and control design}
The following tests examine the use of a frozen QEN after training. Online timing measures query cost under a specified accuracy criterion, and controlled-phase search tests whether its derivatives yield candidates that pass numerical verification.

\subsection{Online queries with one-period numerical reuse}
We compared QEN with numerical propagation that also solves one period and reuses its endpoint and path. For each workload, the QEN's validation p95 of per-member worst-time process infidelity defines an error threshold. We selected the smallest tested numerical resolution satisfying that threshold from a prescribed set of integration step counts. On the RTX 5090, the batch-256 numerical/QEN median time ratios were approximately 2,335 for endpoint queries, 19.7 for 65-node single-period paths, and 73.1 for 24-period grids. These are online timings for the separately identified benchmark endpoint and frozen path, with inputs, weights and analytic reference endpoints already resident on the GPU. Label generation, training and host transfer are excluded. Appendix~\ref{app:cost} specifies the model, fourth-order commutator-free numerical comparator, resolutions, repetitions and accounting scope. The reported errors show that both methods meet the common validation thresholds. The comparison thus measures the cost of producing reusable one-period information within the stated implementation and precision requirements.

\subsection{Controlled-phase design through the frozen QEN}
The QEN can also be differentiated with respect to waveform parameters. We use this property to design the nuclear operation $G_\varphi=\operatorname{diag}(1,1,1,e^{i\varphi})$ at $24T_X=1\,\mu\mathrm{s}$, with fixed device parameters $g_1=g_2=25$ MHz and $\Delta_e=0$. The electron is initialized and projected onto $|0\rangle$. The resulting nuclear block is evaluated with a specified local-$Z$ phase correction, yielding the corrected gate fidelity $F_Z^{\mathrm{fix}}$ defined in Appendix~\ref{app:control}.

For each of eight phases $\varphi=k\pi/4$, $k=1,\ldots,8$, we optimized 64 initial controls through the frozen endpoint network for 500 updates. Projection after each update kept controls within the amplitude and modulation constraints. Every final candidate was evaluated by ODE propagation. The reported acceptance rule requires corrected gate fidelity at least 0.99, population loss at most 0.05 and off-diagonal power at most 0.05. These quantities use the specified electron projection and local-$Z$ correction.

\begin{figure*}[t]\centering
\includegraphics[width=0.90\textwidth]{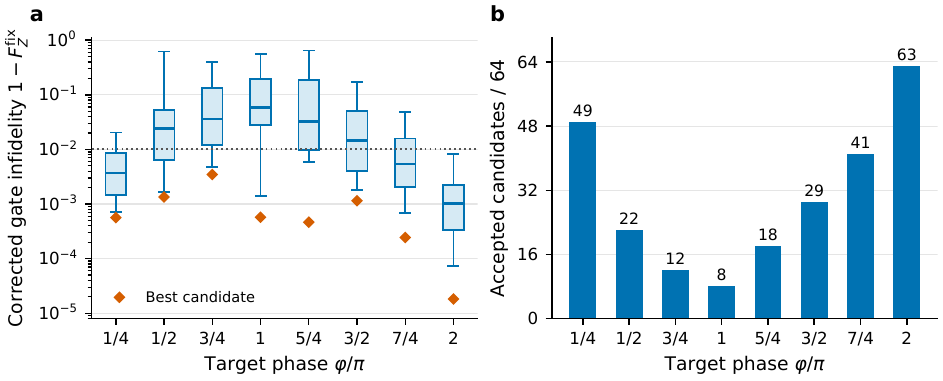}
\caption{ODE-verified controlled-phase candidates generated through the frozen QEN endpoint. (a) Corrected gate infidelity for 64 starts per target phase. Boxes show the interquartile range and median; whiskers span the 2.5th--97.5th percentiles, and diamonds mark the best candidate. The dotted line marks fidelity 0.99. (b) Candidate counts satisfying the joint fidelity, population-loss, and off-diagonal-power criteria. The $2\pi$ target is the identity up to the stated local phase correction.}
\label{fig:control}
\end{figure*}

All eight target phases yielded accepted candidates (Fig.~\ref{fig:control}): between 8 and 63 of 64 starts per phase, with best corrected gate infidelities spanning $1.8\times10^{-5}$ to $3.5\times10^{-3}$ (Table~\ref{tab:control}). Acceptance was hardest at $\varphi=\pi$ and easiest at $2\pi$, while the best candidate at every phase reached fidelity 0.9965 or higher. Across the eight targets, 242 of 512 candidates met the joint criteria. Independent recomputation matched the archived scores within $4.2\times10^{-14}$, with all candidates satisfying the parameter-domain constraints. The same primary endpoint therefore supports forward prediction and gradient-based control search. Numerical verification determines which proposed controls meet the task-specific fidelity, population-loss and off-diagonal-power criteria. This application demonstrates reuse of the trained QEN; a comparison with direct numerical control optimization is outside the present experiment.

\section{Discussion and conclusions}
Quantum Evolution Networks learn complete propagators across prescribed Hamiltonian families through shared neural weights. Their operator output supports predictions for arbitrary input states and makes the approximation accessible to analysis. Unitary decoding and compatible endpoint--path boundaries enforce the stated physical structure; agreement with the target dynamics must be evaluated over the intended parameter and time ranges. Spectral diagnosis, compatible component replacements and worst-input evaluation then inform the use and refinement of the learned model.

For repeated endpoints, single-period process infidelity omits information needed to assess multicycle error. Combining the actual learned residual with finite-time spectral weights in the reference Floquet eigenbasis identifies contributions that accumulate strongly over a chosen horizon, including those associated with near-degenerate gaps outside the exactly commuting sector. Paired direction interventions demonstrate this sensitivity at matched single-period infidelity. These constructed perturbations do not establish that training can produce the same changes. Within the perturbative regime, established coherent-error analysis\cite{chen2026interference} therefore informs repeated use through the measured residual structure and the intended horizon.

Compatible component replacements translate diagnosis into priorities for refinement. In the longitudinal realization, improving the endpoint addresses the dominant error, while further refinement of the already accurate loop offers little benefit at the measured precision. Structured Chebyshev regression also improves the longitudinal endpoint while retaining the learned loop, demonstrating a useful hybrid within this family. For the transverse target window, both endpoint and within-period approximations contribute, with endpoint error increasingly controlling growth at longer times. Refinement priorities thus depend on the physical family and prediction horizon. Since endpoint replacements also update the compatible path boundary, these comparisons identify bottlenecks within the reconstruction without providing an additive decomposition of independent errors.

Reliability must likewise be specified with its conditions of use. The reported empirical reliable horizon combines a worst-input fidelity threshold with required parameter coverage over a sampled time grid. Minimization over input states is exact for the supplied unitaries, while the reported parameter and time coverage is empirical. The coupling-boundary tests show that high coverage within the sampled distribution does not ensure comparable performance at its boundaries. Spectral analysis and direct evaluation also have different scopes: the former estimates leading-order repeated-endpoint process error; the latter evaluates the complete evolution, including within-period motion, against the worst-input criterion.

The continuation experiments identify a further distinction between explaining errors and improving learning. Both ordinary continuation and spectral weighting extend the original empirical reliable horizon, but their ordering varies across the two runs, and the paired parameter-bootstrap intervals do not establish a consistent advantage for spectral weighting. Validation selects the original checkpoint for the direct multicycle objective under the tested protocol. A focused next step is to compare independently trained models at similar single-period accuracy and test whether their residual structures explain differences in multicycle performance. Whether learning targets or representations can consistently suppress the components that accumulate most strongly over a specified horizon remains open.

Repeated simulation queries and control searches\cite{brif2010review,khaneja2005grape,doria2011crab,machnes2018goat} are natural uses of a shared propagator. The online timings quantify query cost after offline label generation and training, leaving total amortized cost to be assessed for a specified workload. In the control study, the frozen QEN supplies derivatives for proposing candidates, and numerical propagation determines which candidates satisfy the gate criteria. These experiments demonstrate reuse within the studied family; they do not establish an advantage over direct numerical control optimization.

The three-spin setting makes this assessment possible with complete reference operators. Extending it to larger registers motivates scalable representations and diagnostics that reduce dependence on those operators. Such extensions must address both the cost of representing the evolution and the cost of assessing its errors, while retaining a meaningful connection between the diagnostic and the desired accuracy criterion.

In conclusion, QEN supports three practical judgments for the studied propagator families: which component should be refined at a chosen horizon, how the measured endpoint residual contributes to leading-order multicycle process error, and over what sampled window the complete prediction meets a specified worst-input fidelity requirement.

\begin{acknowledgments}
This work was supported by the Innovation Program for Quantum Science and Technology (\mbox{2021ZD0301703}), Quantum Science and Technology-National Science and Technology Major Project (Grants No. \mbox{2024ZD0300400}), the Science, Technology and Innovation Commission of Shenzhen Municipality (Grant Nos. \mbox{JCYJ20170412152620376} and \mbox{KYTDPT20181011104202253}), the Shenzhen Science and Technology Program (Grant No. \mbox{KQTD20200820113010023}, and \mbox{ZDYJ20251211121039009}), and CCF-QuantumCtek Superconducting Quantum Computing Special Cooperation Program (No. \mbox{CCF-QC2025002}).
\end{acknowledgments}

\section*{Data availability}
The data supporting the findings of this study are available from the authors upon reasonable request.

\section*{AI-assisted research and manuscript preparation}
AI tools assisted with code development, analysis, figure preparation, and manuscript drafting and editing. The authors are responsible for verification of the computational results and the scientific content.

\appendix
\section{Hamiltonians and unitary network implementations}\label{app:implementations}
This appendix specifies the physical families and the encoders and decoders used in Sec.~\ref{sec:model}. The two realizations share the endpoint--path interface, with coordinates adapted to their available symmetries.

\subsection{Longitudinal Hamiltonian and unitary representation}
We consider an electron spin and two nuclear spins, with Pauli operators $\bm\sigma_e$ and $\bm\sigma_m$ ($m=1,2$). The longitudinal family is
\begin{equation}
H_Z(t)=-\frac{A}{2}\cos(\omega_dt)Z_e-\sum_{m=1}^2\frac{\omega_m}{2}Z_m
+\sum_{m=1}^2\frac{g_m}{4}\bm\sigma_m\cdot\bm\sigma_e.
\label{eq:hz}
\end{equation}
It commutes with $N=\sum_{j=e,1,2}(\Id-Z_j)/2$ at every time. In the computational basis, the eight-dimensional Hilbert space separates into excitation sectors of dimensions $1,3,3,1$. We exploit this exact constraint by predicting a Hermitian endpoint coordinate in a 19-dimensional trace-free block basis, $G_\theta=\sum_{a=1}^{19}b_{\theta,a}B_a$, and applying
\begin{equation}
\widehat\UF=\Cay_Z(G_\theta),\qquad
\Cay_Z(G)=(\Id+iG/2)^{-1}(\Id-iG/2).
\label{eq:cayleyz}
\end{equation}
The map is unitary for Hermitian $G$. The inverse Cayley chart is defined where $I+U$ is invertible. The longitudinal QEN endpoint uses a residual MLP of width 384 and eight residual blocks to map the six physical inputs to endpoint coordinates.

The within-period path removes the analytically known rapid longitudinal rotation,
\begin{align}
U_{\mathrm{fast}}(\tau)&=\exp\!\left[i\frac{A}{2\omega_d}\sin(2\pi\tau)Z_e\right],\\
\widehat V_Z(\tau)&=U_{\mathrm{fast}}(\tau)\widehat R_\phi(\tau)
\Gamma(\tau;\widehat\UF),\qquad
\Gamma(\tau;\UF)=\UF^{\tau}.
\label{eq:zpath}
\end{align}
Fractional powers use the principal eigenphase branch on the retained domain. The residual loop $\widehat R_\phi$ supplies the learned part of the longitudinal within-period path and uses the same block-Cayley construction with coordinates
\begin{equation}
q_a(\tau)=\sum_{k=1}^{24}\left[\alpha_{ak}\sin(2\pi k\tau)
+\beta_{ak}\{1-\cos(2\pi k\tau)\}\right].
\end{equation}
The basis vanishes at both endpoints, so $\widehat R_\phi(0)=\widehat R_\phi(1)=\Id$. An attention encoder with Chebyshev parameter features predicts the coefficients. The residual loop and endpoint connection consequently enforce the path boundaries constructively. Both longitudinal QEN components minimize standardized coordinate errors; the detailed training protocol is given in Appendix~\ref{app:training}.

\subsection{Transverse Hamiltonian and unitary representation}
The transverse family retains the central-spin interactions but adds a carrier-driven electron term,
\begin{equation}
\begin{aligned}
H_X(t)=-\frac{\omega_e}{2}Z_e-\sum_{m=1}^2\frac{\omega_m}{2}Z_m
+\sum_{m=1}^2\frac{g_m}{4}\bm\sigma_m\cdot\bm\sigma_e\\
+A(t)\cos[\omega_ct+\phi(t)]X_e,
\end{aligned}
\label{eq:hx}
\end{equation}
with
\begin{align}
A(t)&=\Omega_0[1+a_1\cos\nu_0t+b_1\sin\nu_0t\notag\\
&\qquad+a_2\cos2\nu_0t+b_2\sin2\nu_0t],\\
\phi(t)&=\beta_1\sin\nu_0t+\beta_2\sin2\nu_0t.
\end{align}
Because the transverse term generally gives $[H_X,N]\neq0$, we use full-space coordinates, which can represent transitions between excitation sectors.

We evaluate this family in the exact carrier frame, $U_X(t)=R_c(t)U_{\mathrm{rot}}(t)$ with $R_c(t)=\exp(i\omega_ctZ_e/2)$. The transformed Hamiltonian retains the counter-rotating drive and all transformed couplings. The fixed carrier-to-modulation ratio is 1340, giving $R_c(T_X)=\Id$. A known frame transformation restores laboratory-frame predictions at intermediate times.

The transverse QEN endpoint decoder starts from the analytic diagonal reference
\begin{equation}
\begin{aligned}
U_{\mathrm{ref}}&=e^{-iT_XH_{\mathrm{diag}}},\qquad\\
H_{\mathrm{diag}}&=-\frac{\Delta_e}{2}Z_e-\sum_m\frac{\omega_m}{2}Z_m
+\sum_m\frac{g_m}{4}Z_mZ_e,
\end{aligned}
\end{equation}
and learns a residual transformation in the full operator space. In the transverse endpoint, a six-block residual MLP with width 376 and a full-feature skip branch maps the 50 scaled parameter features to $3\times63$ real coefficients. With the normalized nonidentity Pauli basis $B_a=P_a/\sqrt8$, the decoder is
\begin{equation}
\begin{aligned}
G_\ell(\lambda)&=\sum_{a=1}^{63}c_{\ell a}(\lambda)B_a,\qquad\\
\widehat\UF&=U_{\mathrm{ref}}\prod_{\ell=1}^{3}\Cay_X(G_\ell),\quad\\
c_{\ell a}&=\frac{\pi}{2}\tanh z_{\ell a},
\end{aligned}
\label{eq:xmlpendpoint}
\end{equation}
where $\Cay_X(G)=(\Id-iG)(\Id+iG)^{-1}$. This supplies operator directions beyond the excitation blocks without prescribing an ordered sequence of individual interaction factors. Three layers with bounded coefficients define the learned residual representation. Z and X thus share coordinate learning and Cayley decoding, while exploiting different available symmetry constraints and using different coordinate conventions.

The endpoint and path are trained separately. At prediction time, the endpoint supplies boundary coordinates to the path decoder, ensuring compatibility of the separately learned components.

For the path, use the normalized full Pauli basis $\{P_a/\sqrt8\}_{a=0}^{63}$ and $\Cay_X(G)=(\Id-iG)(\Id+iG)^{-1}$. Let $\mathbf c_F$ be the Cayley coordinates of $U_{\mathrm{ref}}^\dagger\widehat\UF$. A separate attention network predicts $\mathbf d_\phi(\lambda,\tau)=\tau(1-\tau)\mathbf f_\phi(\lambda,\tau)$, giving
\begin{equation}
\widehat V_{\mathrm{rot}}(\tau)=U_{\mathrm{ref}}(\tau)
\Cay_X\!\left[\sum_{a=0}^{63}\{\tau c_{F,a}+d_{\phi,a}(\lambda,\tau)\}\frac{P_a}{\sqrt8}\right].
\label{eq:xpath}
\end{equation}
Here $U_{\mathrm{ref}}(\tau)=e^{-i\tau T_XH_{\mathrm{diag}}}$, and the laboratory-frame path is $\widehat V_X(\tau)=R_c(\tau T_X)\widehat V_{\mathrm{rot}}(\tau)$. The two boundary values are $\widehat V_X(0)=\Id$ and $\widehat V_X(1)=\widehat\UF$, up to numerical roundoff. The MLP endpoint contains 4,875,583 parameters; the path network contains 2,941,860 parameters. Both transverse QEN components produce unitary factors, but their encoders and decoders differ from the longitudinal realization. The two realizations share the endpoint--path representation and use separately trained parameter encoders.

For the transverse component replacement in Sec.~\ref{sec:xhybrid}, let $U_e$ be the chosen endpoint and $A_e=\Cay_X^{-1}(U_{\mathrm{ref}}^\dagger U_e)$ its Hermitian boundary coordinate. The numerical path residual is $D_{\mathrm{ODE}}(\tau)=\Cay_X^{-1}[U_{\mathrm{ref}}(\tau)^\dagger V_{\mathrm{rot,ODE}}(\tau)]-\tau A_{\mathrm{ODE}}$, and the learned residual is $D_\phi(\tau)=\sum_a d_{\phi,a}(\lambda,\tau)P_a/\sqrt8$. Each hybrid reconstructs $U_{\mathrm{ref}}(\tau)\Cay_X[D(\tau)+\tau A_e]U_e^n$ and restores the common carrier frame. Both residuals vanish at the boundaries. The component replacement keeps the selected residual fixed and updates the boundary coordinate consistently with the chosen endpoint. The comparison uses the frozen primary seed-88001 endpoint, the frozen path network, and the degree-three regression endpoint described in Sec.~\ref{sec:qicomparison}. The 256 members and 81 phases match the primary spectral subset; numerical paths use the same 4,096/8,192-step Richardson--polar reference. Independent 8,192/16,384-step refinement on eight members gave a maximum reference-to-reference process infidelity of $1.1\times10^{-10}$ during the final period of the 24-period window. The unmodified QEN reconstruction reproduced the archived worst-time process scores within $3.4\times10^{-13}$.

\begin{table*}[t]\centering\small
\caption{Network components and learned output representations. Width and block count refer to the parameter encoders.}\label{tab:architecture}
\begin{tabular}{llll}\toprule
Branch & Encoder & Width / blocks & Learned output\\\midrule
Z endpoint & Residual MLP & 384 / 8 & 19 block coordinates\\
Z path & Chebyshev features, attention & 128 / 4 & 24 harmonic pairs per coordinate\\
X endpoint & Residual MLP + feature skip & 376 / 6 & $3\times63$ Cayley coefficients\\
X path & Parameter/time attention & 256 / 4 & 64 residual coordinates\\\bottomrule
\end{tabular}
\end{table*}

\section{Models and evaluation scopes}\label{app:modelmap}
Model identity is determined by weights and selection history, while each result specifies its own reference and query scope. The original seed-88001 endpoint is the same MLP checkpoint used in the shared-Cayley encoder comparison and in the original O branch of the continuation study. Encoder-study scores refer to repeated endpoints; the main result evaluates the combined endpoint and path at all queried times. The QEN endpoint in the FNO comparison is a separate training run. The numerical query benchmark uses the spectral-weighted endpoint selected for that benchmark, with the same path architecture. These identities are summarized in Table~\ref{tab:modelmap}. The encoder study uses the original data payload's 2,048-member test split and its stored endpoint references. The continuation endpoint comparison uses a separate 2,048-member common test set, with no parameter members shared with that split, and newly computed 4,096/8,192-step Richardson--polar references. Consequently, the same original weights yield different sample percentiles in Tables~\ref{tab:cayleyencoders} and~\ref{tab:continuedendpoint}; both tables report repeated endpoints.
\begin{table*}[t]\centering\small
\caption{Model-to-result correspondence. The main path is the frozen four-layer attention path defined in Appendix~\ref{app:training}. Integer-time evaluations use only the endpoint. Common-test denotes the repeatedly evaluated fixed parameter set described in Sec.~\ref{sec:protocol}.}\label{tab:modelmap}
\begin{tabularx}{\linewidth}{@{}p{.19\linewidth}XXX@{}}\toprule
Result & Endpoint and selection & Path & Evaluation\\\midrule
Main accuracy and control & Original MLP, seed 88001; validation ranks repeated-endpoint fidelity & Main path for dynamics; endpoint only for control & 2,048 process / 256 spectral members; 512 controls\\
FNO comparison & Separate QEN run, seed 88001, update 31,000; FNO update 13,000 & Endpoint only & 2,048 comparison test parameters\\
Encoder comparison & Attention and MLP, seeds 88001/88002; common validation rule & Endpoint only & 2,048 encoder-study test parameters\\
Continuation & O and A/B/C for both seeds; validation $H^*$ then tail fidelity & Main path, with each predicted endpoint & 2,048 common-test endpoints / 256 members over the full time grid\\
Online timing & Benchmark B endpoint, seed 88001 & Main path for path queries & Validation tolerance; batch 1 and 256\\\bottomrule
\end{tabularx}
\end{table*}

\section{Training protocols}\label{app:training}
Longitudinal coordinate labels follow a trace-aligned phase convention. For an ODE endpoint $U$, set $\widetilde U=e^{-i\arg\Tr U}U$, form $G_0=2i(\widetilde U+I)^{-1}(\widetilde U-I)$, and Hermitian-symmetrize it. The training target is the trace-free projection $G=G_0-\Tr(G_0)I/8$, expanded in the 19-dimensional block basis. The decoded prediction is compared with the ODE operator using phase-invariant fidelity. This convention fixes both the global-phase alignment and the coordinate projection.

The longitudinal endpoint is a width-384 residual MLP with eight pre-normalized blocks, SiLU activations, and 19 output coordinates, containing 2,381,587 parameters. Training minimizes standardized coordinate MSE over 38,400 AdamW updates\cite{loshchilov2019adamw}, with batch size 256, initial learning rate $10^{-4}$, weight decay $10^{-7}$, gradient clipping at one, and cosine decay. Validation runs every 640 updates. Checkpoints are selected lexicographically by the first percentile of repeated-endpoint $F_{1000}$, its fifth percentile, and the fifth percentile of $F_{50}$.

The longitudinal loop uses degree-12 Chebyshev parameter features and four pre-normalized attention layers\cite{vaswani2017attention}, with width 128 and eight heads. Its output comprises 24 sine and shifted-cosine harmonics per coordinate. Training uses standardized coordinate MSE, batch size 64, initial learning rate $10^{-4}$, weight decay $10^{-7}$, and at most 300 epochs, with validation-based early stopping after epoch 150. Normalization statistics are computed on the training data.

The transverse input features are built from a scaled ten-component vector $x$. Its first four entries are $(g_1-25)/15$, $(g_2-25)/15$, $\Delta_e/2$, and $(\Omega_0-11)/3$, with frequency inputs in MHz. The remaining entries divide the amplitude and phase coefficients by their respective ball radii. Concatenating $x$, $\sin(\pi x)$, $\cos(\pi x)$, $\sin(2\pi x)$, and $\cos(2\pi x)$ gives 50 features, standardized using training-set statistics.

The transverse endpoint has six residual MLP blocks, width 376, feed-forward width 1,024, and a full-feature skip branch. Three Cayley layers output 189 coefficients, giving 4,875,583 trainable parameters. Each block applies layer normalization and a two-layer SiLU feed-forward transformation with residual addition. The output head receives the final normalized hidden vector concatenated with the 50 input features. The loss is the squared Frobenius norm of the inverse-Cayley coordinate of the relative endpoint $\UF^\dagger\widehat\UF$, divided by eight. Training uses batch size 256, AdamW, zero weight decay, and gradient clipping at one. A first stage of 30,000 updates at learning rate $10^{-4}$ is followed by 2,000 updates at $3\times10^{-5}$, initialized from the selected first-stage checkpoint with reset optimizer and sampling generator. Both stages use the single-period loss. Selection ranks validation repeated-endpoint $F_{100}$ p05 followed by $F_{24}$ p05, evaluated every 1,000 updates and at the first update of each stage. Of the two MLP seeds, 88001 supplies the endpoint for evaluation over the full time grid, selected by its higher validation score.

The transverse path uses parameter and time tokens, four attention layers, width 256, four heads, and feed-forward width 768, totaling 2,941,860 parameters. It minimizes MSE in the 64-coordinate path residual over 12,000 AdamW updates at learning rate $10^{-4}$, with zero weight decay and unit gradient clipping. Its training pool contains 16,384 parameters partitioned into 32-member shards. A shard is sampled with replacement every 50 updates, and each batch draws 256 parameter--time pairs from that shard. Training samples internal nodes of the 8,192-point path. Checkpoint selection evaluates the first 512 validation members every 2,000 updates, using one-period worst-time process p05 with the reference endpoint. In prediction over the full time grid, the learned endpoint supplies both the repeated-period factor and the path boundary coordinates.

The path's nine time features are $\tau$, $\tau(1-\tau)$, sine and cosine pairs at $2\pi k\tau$ for $k=1,2,4$, and $\sin(2\pi\times1340\tau)$. Linear projections form a parameter token and a time token. Their pooled attention output is concatenated with the parameter and time features before the 64-coordinate head.

\section{Reference propagation and interface checks}\label{app:reference}
Longitudinal labels are generated by complex128 midpoint propagation with 2,560 steps per period, supplemented by independent adaptive ODE checks on prescribed subsets. The longitudinal time grid evaluates 65 within-period nodes, including both boundaries, for period indices $n=0,\ldots,1000$. Its endpoint is $1001T_Z$, spanning approximately 1.82--2.47 $\mu$s as the drive frequency varies. The 2,048-member set includes predefined sampling strata, and its reported quantiles summarize those members.

Transverse endpoint training labels use 8,192/16,384-step fourth-order commutator-free propagation, while path training uses an 8,192-node path. Evaluation over the full time grid uses 4,096/8,192-step propagation. For step counts $N$ and $2N$, Richardson extrapolation constructs $(16U_{2N}-U_N)/15$, followed by polar projection. All requested local times are aligned to the integration grid. The 65-node process grid is augmented by 16 phases at odd multiples of $1/2048$ for the spectral subset. The first eight test members are additionally evaluated with 8,192/16,384 steps; the maximum reference-to-reference half diamond distance on the check grid through 101 periods is $6.9\times10^{-5}$.

On these interface-check members, the initial operator equals the identity to the recorded precision, the largest endpoint mismatch is $1.2\times10^{-15}$, and the largest path unitarity defect is $2.3\times10^{-15}$. The actual inverse-Cayley input also satisfies
\begin{equation}
\min_\lambda\sigma_{\min}(I+U_{\mathrm{ref}}^\dagger\widehat\UF)=1.36
\end{equation}
on this check set. This probes the conditioning of the boundary-coordinate transformation used by the path decoder. These interface checks establish the numerical consistency needed to interpret the component replacements.

\section{Spectral evaluation and process-fidelity bound}\label{app:bound}
For unitary operators, process fidelity supplies the conservative worst-input bound
\begin{equation}
\Fw\geq\left[\max\{0,1-d(1-\sqrt{\Fpro})\}\right]^2.
\label{eq:bound}
\end{equation}
The following derivation also establishes its use for worst-time scores.
Let $W=U^\dagger\widehat U$ be unitary, and rotate its global phase so that $\Tr W=d\sqrt{\Fpro}$ is real and nonnegative. For eigenphases $\theta_j$,
\begin{equation}
\sum_j(1-\cos\theta_j)=d(1-\sqrt{\Fpro})\equiv S.
\end{equation}
Each summand is nonnegative, so $\operatorname{Re}\langle\psi|W|\psi\rangle\geq1-S$ for every normalized state. Squaring the positive part yields Eq.~\eqref{eq:bound}. By monotonicity, the bound also applies to worst-time process scores. For the longitudinal set evaluated over the full time grid, the resulting worst-input lower-bound p05 is 0.9922, and 96.4\% of members have a lower bound of at least 0.99.

The direct transverse calculation sorts the eigenphases of $W$ and finds their largest circular gap. The complementary arc is the shortest containing arc, which determines the distance from the origin to the spectral convex hull. Analytic checks cover identity operators, common phases, two-extreme-phase spectra, and spectra whose convex hull contains the origin. Numerical checks compare this formula with random-state overlaps and verify the ordering between process fidelity, worst-input fidelity, and the conservative lower bound. Cumulative scores are formed by taking minima as the evaluated time window grows.

\section{Longitudinal representation comparisons}\label{app:zcomparisons}
\begin{figure*}[t]\centering
\includegraphics[width=0.90\textwidth]{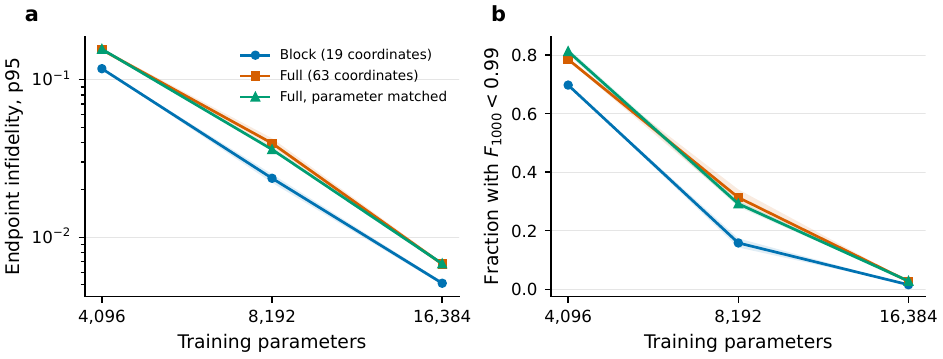}
\caption{Longitudinal endpoint representation comparison after 1,000 repetitions. (a) The 95th-percentile endpoint infidelity. (b) The fraction of test parameters with fidelity below 0.99. Curves average two training seeds; shadows span their values. The full-space comparisons use the same-width and approximately parameter-matched encoders.}
\label{fig:symmetry}
\end{figure*}

Figure~\ref{fig:symmetry} compares block coordinates with full-space coordinates at three training sizes. The approximately parameter-matched pair contains 799,251 and 798,131 parameters, with seeds 1002 and 1003. Both models use standardized coordinate losses. In the full-space representation, normalization uses a $10^{-8}$ scale floor, and the loss averages all 63 channels, including symmetry-forbidden target coordinates. The comparison therefore evaluates the complete representation and weighting choices used by each model.

\begin{figure*}[t]\centering
\includegraphics[width=0.84\textwidth]{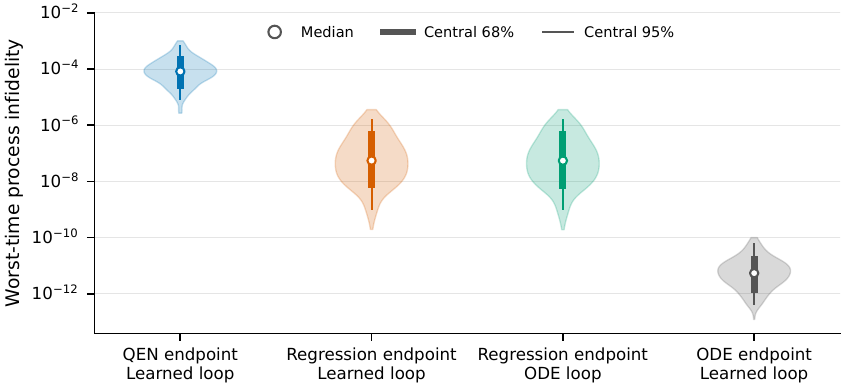}
\caption{Longitudinal hybrid comparison on a common 256-member subset. Each member contributes its maximum process infidelity over the evaluated time grid. Shadows show the distribution of test-set data points. Open circles mark medians, thick lines span the 16th--84th percentiles, and thin lines span the 2.5th--97.5th percentiles. These ranges summarize variation across parameter members. All errors are evaluated relative to the fixed ODE reference.}
\label{fig:hybrid}
\end{figure*}

The Chebyshev endpoint uses degree-12 unary and pairwise Chebyshev features and degree-3 triple interactions. Combined with the learned loop, it yields process-infidelity p95 of $1.3\times10^{-6}$ on the common 256-member subset. The neural endpoint with the same loop gives $4.1\times10^{-4}$. The distributions in Fig.~\ref{fig:hybrid} show the dependence on endpoint choice and the close agreement between learned-loop and ODE-loop reconstructions for a fixed regression endpoint.

\section{Transverse channel-error distributions}\label{app:channel}
The Haar-averaged pure-state fidelity is $F_{\mathrm{avg}}=(d\Fpro+1)/(d+1)$\cite{nielsen2002fidelity}. For the associated unitary channels, the half diamond distance is
\begin{equation}
D_\diamond(\mathcal U,\widehat{\mathcal U})
=\tfrac12\|\mathcal U-\widehat{\mathcal U}\|_\diamond
=\sqrt{1-\Fw}.
\label{eq:diamond}
\end{equation}
It quantifies channel distinguishability, including auxiliary systems\cite{haah2023unitary}, and is a monotone conversion of the worst-input error.

\begin{figure*}[t]\centering
\includegraphics[width=0.90\textwidth]{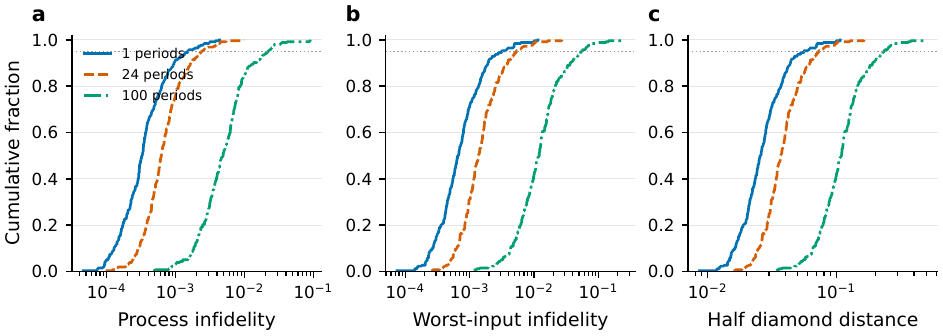}
\caption{Transverse error distributions on the common 256-member spectral subset. Each member contributes its largest error over the specified time grid. (a) Process infidelity, (b) worst-input infidelity, and (c) half diamond distance. Dotted horizontal lines mark cumulative fraction 0.95. The three line styles identify the one-, 24-, and 100-period windows.}
\label{fig:channel}
\end{figure*}

Figure~\ref{fig:channel} evaluates all three error measures on the same 256 members and augmented time grid. At $T_X$, $24T_X$, and $100T_X$, the worst-input fidelity p05 is 0.9969, 0.9947, and 0.9460, and the corresponding half diamond distance p95 is 0.0559, 0.0728, and 0.2324. Reporting both quantities relates the overlap criterion to an operational channel-distance tolerance. The pointwise bootstrap bands in Fig.~\ref{fig:worst} use 10,000 resamples of 256 members, with seed 280926.

\section{Controlled-phase objective and evaluation}\label{app:control}
With basis order $|e,n_1,n_2\rangle$, we project the electron onto $|0\rangle$ after evolution from the same electron state. The resulting nuclear block is $K=\langle0_e|U(24T_X)|0_e\rangle$. For target $G_\varphi$, the average unnormalized state overlap is
\begin{equation}
F_{\mathrm{strict}}=\frac{\Tr(K^\dagger K)+|\Tr(G_\varphi^\dagger K)|^2}{20}.
\end{equation}
Population loss and off-diagonal power are
\begin{equation}
L=1-\frac{\Tr(K^\dagger K)}4,\qquad O=\frac14\sum_{j\ne k}|K_{jk}|^2.
\end{equation}
Write $K_{jj}=a_je^{i\theta_j}$ and define the conditional phase $\Theta=\operatorname{wrap}(\theta_0-\theta_1-\theta_2+\theta_3)$. A fixed global and local-$Z$ correction aligns the first three diagonal phases. The remaining target mismatch is $\delta=\operatorname{wrap}(\Theta-\varphi)$, giving
\begin{equation}
F_Z^{\mathrm{fix}}=\frac{\Tr(K^\dagger K)+|a_0+a_1+a_2+a_3e^{i\delta}|^2}{20}.
\end{equation}
This convention evaluates the gate with tracked local phase corrections and retains the contribution of electron population loss in the score.

The surrogate objective is
\begin{equation}
\mathcal J=\operatorname{wrap}(\Theta-\varphi)^2+10L^2+10O+10^{-4}P_{\mathrm{amp}},
\end{equation}
where $P_{\mathrm{amp}}$ penalizes squared violations of the amplitude interval and modulation constraints. Each target uses 64 starts with seed 61001, 500 Adam updates, and learning rate 0.03. After each update, controls are projected into the admissible domain. All final 512 candidates satisfy the prescribed control constraints to numerical precision. Final evaluation uses the 4,096/8,192-step Richardson--polar ODE endpoint raised to the 24th power. Table~\ref{tab:control} summarizes the results.

\begin{table}[t]\centering\small
\caption{ODE-verified controlled-phase results for the frozen QEN. Acceptance requires $F_Z^{\mathrm{fix}}\geq0.99$, $L\leq0.05$, and $O\leq0.05$.}\label{tab:control}
\begin{tabular}{rrrr}\toprule
$\varphi/\pi$ & Accepted / 64 & Best $F_Z^{\mathrm{fix}}$ & Median $F_Z^{\mathrm{fix}}$\\\midrule
1/4 & 49 & 0.9994 & 0.9963\\
1/2 & 22 & 0.9987 & 0.9756\\
3/4 & 12 & 0.9965 & 0.9638\\
1 & 8 & 0.9994 & 0.9416\\
5/4 & 18 & 0.9995 & 0.9672\\
3/2 & 29 & 0.9988 & 0.9852\\
7/4 & 41 & 0.9998 & 0.9946\\
2 & 63 & ${>}0.9999$ & 0.9990\\\bottomrule
\end{tabular}
\end{table}

\section{Trajectory evaluation}\label{app:trajectories}
\begin{figure*}[t]\centering
\includegraphics[width=0.90\textwidth]{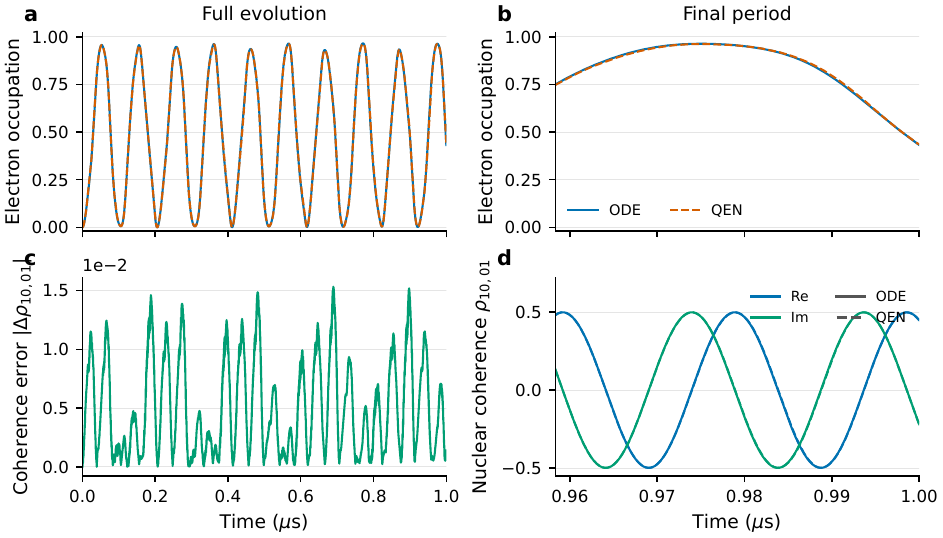}
\caption{Observable dynamics generated by one complete QEN prediction for a fixed held-out parameter member and initial state $(|010\rangle+i|001\rangle)/\sqrt2$. (a,b) Electron occupation over the full window and its final period. (c) Absolute error of the complex reduced nuclear coherence over the full window, peaking at $1.5\times10^{-2}$, about 3.1\% of the 0.5 coherence amplitude. (d) Real and imaginary nuclear coherence during the final period; solid and dashed lines denote ODE and QEN predictions. The displayed member (index 1,524) was fixed before this evaluation. The dense grid contains 6,145 times through $24T_X$.}
\label{fig:dynamics}
\end{figure*}

Two illustrative members, with zero-based indices 1524 and 1569 in the 2,048-member parameter array, were fixed before evaluating the MLP endpoint. Saved trajectories include both $|100\rangle$ and $(|010\rangle+i|001\rangle)/\sqrt2$ as initial states. Figure~\ref{fig:dynamics} uses the coherent input of member 1524. Reference and predicted operators are evaluated at 257 within-period nodes, with duplicate boundaries removed, producing 6,145 times through $24T_X$. Reference paths use 4,096/8,192-step Richardson--polar propagation. The maximum Frobenius unitarity defect of the saved predictions is $4.7\times10^{-14}$. Electron occupation is $\langle(\Id-Z_e)/2\rangle$, and nuclear coherence is obtained after tracing out the electron.

Figure~\ref{fig:dynamics} illustrates the learned operator's action on the coherent nuclear input $(|010\rangle+i|001\rangle)/\sqrt2$, using basis order $|e,n_1,n_2\rangle$. Electron occupation tracks population transfer, while the complex reduced nuclear coherence $\rho_{10,01}$ probes relative-phase dynamics. The QEN follows the ODE trajectory over the full microsecond window and resolves the oscillations within its final period. The coherence-error trace over the full time grid displays their difference separately. This example illustrates observable dynamics; the worst-input evaluation in Sec.~\ref{sec:mainresults} assesses accuracy over all initial states.

\begin{figure*}[t]\centering
\includegraphics[width=0.94\linewidth]{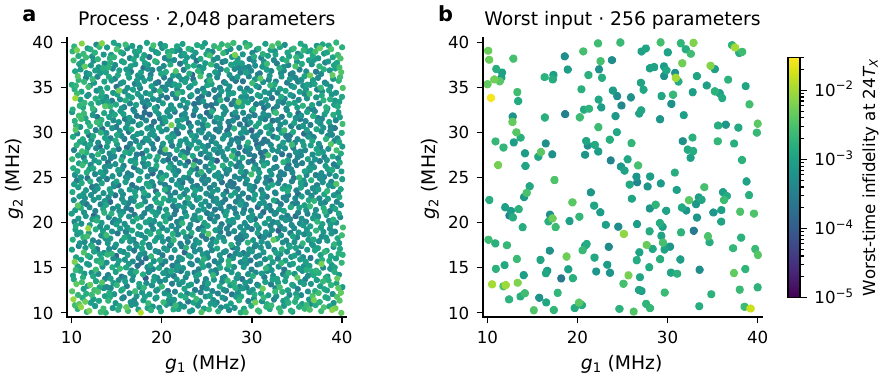}
\caption{Parameter-resolved accuracy over $24T_X$. Colors show worst-time process infidelity for the full 2,048-member set (a), and worst-input infidelity for the 256-member spectral subset (b). Both panels use the same logarithmic color scale. The points project the ten-dimensional parameter sample onto the two electron--nuclear couplings.}
\label{fig:parammap}
\end{figure*}

The parameter projection in Fig.~\ref{fig:parammap} places complete-operator errors in the coupling plane. Each point retains its independently sampled values of the other eight inputs. The view locates high-fidelity predictions and larger errors within the tested sample, complementing the aggregate quantiles. Both panels summarize the same physical window, with the spectral panel resolving the most sensitive input state for each member.

\section{Transverse encoder and decoder comparisons}\label{app:xablation}
\subsection{Output representation at a fixed update budget}
The first comparison used 16,384 training, 2,048 validation, and 2,048 test members. Models shared 50 input features, training normalization, the analytic diagonal reference, single-period relative-Cayley loss, batch size 256, AdamW learning rate $10^{-4}$, zero weight decay, and unit gradient clipping. Training lasted 12,000 updates per arm. Validation was evaluated every 1,000 updates, and checkpoints were ranked by repeated-endpoint $F_{100}$ p05, then $F_{24}$ p05, then $F_1$ p05.

The attention and parameter-matched MLP encoders had 4,853,124 and 4,854,217 parameters with the physical-factor decoder. This decoder contains six layers, each with 18 coefficients for diagonal, electron--nuclear, transverse-electron, and global-phase factors. The alternative uses three full-space Cayley factors as in Eq.~\eqref{eq:xmlpendpoint}. Each decoder therefore defines its own operator basis, factor count, and coefficient range, while the analytic reference is shared. Four independently initialized seeds provide the comparisons in Table~\ref{tab:shortablations}.

\begin{table*}[t]\centering\small
\caption{Selected validation repeated-endpoint $F_{100}$ p05 after 12,000 updates. Columns identify independent training seeds.}\label{tab:shortablations}
\begin{tabular}{lrrrr}\toprule
Encoder / decoder & 88001 & 88002 & 88003 & 88004\\\midrule
Attention / physical factors & 0.7263 & 0.7572 & 0.6424 & 0.7777\\
Matched MLP / physical factors & 0.5836 & 0.7981 & 0.7351 & 0.8076\\
Attention / Cayley $\times3$ & 0.8412 & 0.8039 & 0.7963 & 0.8829\\\bottomrule
\end{tabular}
\end{table*}

The three mean selected scores were 0.7259, 0.7311, and 0.8311, respectively. The full-space Cayley decoder improved the attention model's score in all four seeds, by 0.047--0.154. Encoder ordering varied across seeds. These observations motivated the longer shared-Cayley comparison below.

\subsection{Encoder accuracy and training cost with a shared Cayley decoder}
The longer comparison fixed the three-layer Cayley decoder. Attention used six layers, width 256, eight heads, and feed-forward width 1,024; the residual MLP used six blocks, width 376, and the same feed-forward width. Both retained the feature skip branch. Training and validation selection followed Appendix~\ref{app:training}, using 30,000 initial and 2,000 continuation updates. The selected checkpoint was loaded before held-out testing.

\begin{table}[t]\centering\small
\caption{Encoder comparison with shared three-layer Cayley decoding. Fidelities are parameter p05 values for the repeated endpoint $\widehat\UF^h$ on 2,048 held-out members. Training time is synchronized cumulative update time for all 32,000 updates on an RTX 4060 Laptop GPU; validation and file operations are accounted separately. The attention and MLP models contain 4,877,991 and 4,875,583 parameters.}\label{tab:cayleyencoders}
\begin{tabular}{llrrr}\toprule
Encoder & Seed & $F_{24}$ p05 & $F_{100}$ p05 & Training (min)\\\midrule
Attention & 88001 & 0.9985 & 0.9752 & 99.9\\
Residual MLP & 88001 & 0.9987 & 0.9795 & 31.7\\
Attention & 88002 & 0.9988 & 0.9816 & 101.0\\
Residual MLP & 88002 & 0.9984 & 0.9740 & 34.0\\\bottomrule
\end{tabular}
\end{table}

All inputs, labels, split sizes, batch sizes, and normalization were shared. Computation used float64/complex128 with PyTorch 2.7.1. Run order was MLP then attention for seed 88001 and attention then MLP for seed 88002. Synchronization bracketed each update, and the timing includes updates after the selected checkpoint. Total update times were 200.9 minutes for attention and 65.7 minutes for the MLP, a factor of 3.06 under this protocol. The encoder accuracy ordering reversed between the two seeds (Table~\ref{tab:cayleyencoders}).

\section{FNO comparison protocol}\label{app:qicomparison}
The comparator follows the Floquet-Hamiltonian learning target of Qi et al.\cite{qi2026floquetfno}. For this register, the implementation represents input Hamiltonians and output generators in the complete traceless Pauli basis, uses the aligned 65-node period grid, and reconstructs the endpoint at the physical starting phase zero. Basis size, grid, width, mode count, optimizer schedule and validation criterion are specified here as the configuration of the adapted experiment.

The comparison used 16,384 training, 2,048 validation, and 2,048 test configurations from the transverse endpoint dataset, with training seed 88001. The Floquet-Hamiltonian FNO used width 256, three Fourier layers, 32 retained low-frequency modes, GELU activations, and 6,587,199 trainable parameters. Its outputs were coefficients of $T_X H_{\mathrm F}(t)$ in the 63-dimensional traceless Pauli basis. Training minimized raw coefficient mean-squared error with batch size 256 for 12,000 updates at learning rate $10^{-4}$ followed by 2,000 at $3\times10^{-5}$. Validation coefficient error selected update 13,000, with MSE $7.56\times10^{-3}$. This run used an NVIDIA RTX 4060 Laptop GPU.

Hamiltonian inputs and generator targets shared the 65-node grid $t_j=jT_X/64$, $j=0,\ldots,64$. The conjugation path used 1,024 fourth-order commutator-free integration steps per period, with 16 steps between output nodes. The effective generator at zero was obtained from the high-resolution reference endpoint, and the remaining targets were constructed as $H_{\mathrm F}(t_j)=P(t_j)H_{\mathrm F}(0)P(t_j)^\dagger$. Table~\ref{tab:qicomparison} evaluates the fixed starting-phase output at zero and its integer powers against the high-resolution reference endpoints.

The QEN comparison run used the residual MLP and three-factor full-space Cayley decoder with 4,875,583 parameters. It used 30,000 updates at learning rate $10^{-4}$ and 2,000 continuation updates at $3\times10^{-5}$; validation selection chose update 31,000. This is the seed-88001 endpoint comparison run, distinct from the seed-88001 encoder study in Table~\ref{tab:cayleyencoders}. These comparisons evaluate the two trained implementations with their respective supervision and training schedules. Process fidelity is computed between the complete predicted and reference operators, with the same starting phase and period count for both methods.

\section{Spectral error diagnosis and continuation protocol}\label{app:spectral}
For the spectrum-averaged comparison in Sec.~\ref{sec:spectral},
\begin{equation}
\begin{aligned}
P_1(n)&=c_n[1-\Fpro(U,\widehat U)],\\
c_n&=\frac{\sum_{a,b}w_n(\Delta_{ab})-n^2}{d^2-1}.
\end{aligned}
\end{equation}
This estimate averages over orientations of a traceless residual at the fixed reference spectrum, while $P_2$ retains the measured residual orientation.

With the identity component of $E$ removed, the relation $\widehat U=Ue^{-iE}$ holds up to a global phase. The first-order accumulated error is the conjugation sum $E_n=\sum_{k=0}^{n-1}U^{-k}EU^k$. For the traceless accumulated generator, the leading infidelity is $1-\Fpro\simeq\Tr(E_n^2)/d$. The sum yields Eq.~\eqref{eq:spectralprediction}; for a nonzero eigenphase difference, its weight can equivalently be written as $\sin^2(n\Delta/2)/\sin^2(\Delta/2)$, with continuous limit $n^2$ at zero. Spectral differences are evaluated on the phase circle. Predictions are compared to the exact powers, and the region with accumulated generator spectral width at most 0.3 rad is recorded separately in the numerical analysis. This tests the perturbative estimate against repeated endpoints under the same numerical reference.

For residual-orientation interventions, the commuting projection keeps diagonal entries and complete eigenspace blocks whose circular eigenphase gap is below $10^{-10}$ rad. The complementary projection retains the other entries. Each projected generator is rescaled by solving for the original single-period process infidelity. Members are jointly admitted when both directions have a successful fidelity match with rescaling at most 10. In the plotted original-model subsets, the largest matching residual is $4.4\times10^{-16}$, each retained component has at least 0.10 of the original generator norm, and the bound $2\|E_{\rm scaled}\|_F$ on its spectral width is at most 0.064 rad. The same 254 and 255 members are used for both directions in the two panels. Exact powers evaluate both directions through 100 periods. These paired perturbations describe sensitivity of the frozen learned residual.

Continuation uses AdamW with zero weight decay, learning rate $3\times10^{-5}$, batch size 256, gradient clipping at 1, and 2,000 updates. The three objectives share initial weights, optimizer resets and batch draws. Let $G=g\Id+G_0$ be the relative Cayley coordinate, $f=\|G_0\|_F^2/d$, and $\overline w$ the spectral weights averaged over $\{24,50,100\}$. A minimizes the batch mean of $g^2+f$. B minimizes $g^2+(1-\beta)f+\beta\sum_{ab}\overline w_{ab}|(Q^\dagger G_0Q)_{ab}|^2/(dc)$, where $\beta=0.5$ and the training-set normalization is $c=520.33$. C replaces the weighted term by the mean exact multicycle process infidelity divided by $\kappa^2c$, with $\kappa\simeq2$ for the Cayley convention used. Training-set constants are frozen before continuation.

Validation uses a fixed 256-member sample and cumulative integer-period worst-input fidelity through 100 periods. Checkpoints at updates 0, 250, \ldots, 2,000 are ordered by $H^*$, then p05 at 100 periods, p05 at 24 periods, and earlier update. The coverage rule requires at least 244 of 256 members above fidelity 0.99. Subsequent common-test endpoint evaluation uses 2,048 parameters, while evaluation over the full time grid uses the fixed 256-member subset and 81 within-period nodes. Each predicted endpoint is inserted into the frozen path network's boundary condition.

Paired uncertainty intervals resample parameter members 10,000 times with seed 290930, sharing each draw between the compared models. These intervals describe parameter-sample variation for the frozen checkpoints. At 100 periods, the B-minus-A p05 difference for seed 88001 is $-0.0019$, with a 95\% interval of $[-0.0102,0.0070]$. All comparison classes are computed for each seed before interpretation. The complete 2,048-member repeated-endpoint comparison is listed in Table~\ref{tab:continuedendpoint}. Additional checkpoint comparisons match validation single-period infidelity p95 within 10\% before evaluating their common-test long-time behavior.

\begin{table}[t]\centering\footnotesize\setlength{\tabcolsep}{3pt}
\caption{Repeated-endpoint process fidelity p05 for all validation-selected endpoints on the common 2,048-member test set. The parameter members differ from those in Table~\ref{tab:cayleyencoders}, with zero overlap; references are recomputed using 4,096/8,192-step Richardson--polar propagation. These integer-time scores are distinct from the worst-time worst-input scores in Table~\ref{tab:spectraltraining}.}\label{tab:continuedendpoint}
\begin{tabular}{llrrrr}\toprule
Seed & Objective & $F_1$ & $F_{24}$ & $F_{50}$ & $F_{100}$\\\midrule
88001 & O & $1-1.7\times10^{-5}$ & 0.9988 & 0.9948 & 0.9796\\
88001 & A & $1-1.4\times10^{-5}$ & 0.9989 & 0.9956 & 0.9828\\
88001 & B & $1-1.5\times10^{-5}$ & 0.9988 & 0.9951 & 0.9807\\
88001 & C & $1-1.7\times10^{-5}$ & 0.9988 & 0.9948 & 0.9796\\
88002 & O & $1-2.2\times10^{-5}$ & 0.9985 & 0.9937 & 0.9751\\
88002 & A & $1-2.0\times10^{-5}$ & 0.9986 & 0.9941 & 0.9768\\
88002 & B & $1-1.9\times10^{-5}$ & 0.9987 & 0.9946 & 0.9789\\
88002 & C & $1-2.2\times10^{-5}$ & 0.9985 & 0.9937 & 0.9751\\
\bottomrule\end{tabular}\end{table}

\section{Reference sensitivity and coordinate diagnostics}\label{app:newreference}
Reference refinement compares 4,096/8,192-step and 8,192/16,384-step extrapolated, polar-projected propagators on 32 preselected members: eight random parameters, eight original-model tail members, eight coupling-boundary members and eight current control candidates. Queries cover within-period nodes, 24 and 100 periods, and the neighborhood of each final model's reliable horizon. The maximum change in reported worst-input fidelity is $3.0\times10^{-6}$ at 24 periods and $4.2\times10^{-5}$ at 100 periods. Threshold decisions at 0.99 and the subset reliable horizons are unchanged in this refinement.

The 72-member boundary sample consists of eight interior anchors and, per anchor, four coupling-face points and four coupling-corner points at the endpoints of $g_1,g_2\in[10,40]$ MHz. Other inputs retain their anchor values. These samples are evaluated separately from the common-test distribution. For the original seed-88001 model, all eight anchors, 27 of 32 face members and 11 of 32 corner members retain worst-time worst-input fidelity at least 0.99 through 24 periods. The corresponding p05 scores are 0.9918, 0.9831 and 0.9343. This stress test distinguishes distribution-sampled accuracy from coverage of the coupling-box boundary. Coordinate conditioning uses the actual matrices $I+U_{\rm ref}^\dagger\widehat U$ for path decoding and $I+U^\dagger\widehat U$ for the relative training coordinate. The common-test, boundary, and current control samples are finite throughout the evaluated models; for the control sample the smallest singular value in the reference-residual channel is about 1.22 and the largest condition number about 1.59.

The FNO comparator uses the 1,024-step conjugation-path labels specified in Appendix~\ref{app:qicomparison}. An independent 32-member, 65-node audit compares these paths with 4,096-step integration and the high-resolution endpoint reference. Median endpoint process infidelity is $1.6\times10^{-5}$ at 1,024 steps and $4.0\times10^{-9}$ at 4,096 steps; trace-free coordinate encoding and decoding has median Frobenius residual $3.3\times10^{-12}$. Integration accuracy and coordinate round-trip accuracy are evaluated separately.

\section{Numerical queries under a common error threshold}\label{app:cost}
The numerical comparator and QEN both compute one-period information and reuse it. The three workloads are an endpoint query, a 65-node single-period path, and that path reused over 24 periods. For each workload, validation fixes a process-infidelity tolerance by taking the maximum error over queried times for each member and then the parameter p95. Numerical resolution is the smallest tested choice satisfying this threshold among 256, 512, 1,024, 2,048, 4,096 and 8,192 integration steps. The selected resolutions are 2,048, 512 and 2,048 steps, respectively; their measured errors and the QEN thresholds are listed in Table~\ref{tab:matchingprecision}. The discrete resolution choices produce numerical errors below the thresholds, rather than identical errors for both methods. The timing uses the spectral-weighted seed-88001 endpoint evaluated in this benchmark and the frozen path network.

On an RTX 5090, with float64/complex128 arithmetic, two warmups and five synchronized repetitions, the median numerical/QEN time ratios are 3,196, 217 and 780 for batch 1, and 2,335, 19.7 and 73.1 for batch 256 (Table~\ref{tab:onlinecost}). Outputs and repeated matrix products remain on the GPU. Inputs, learned weights and analytic reference endpoints are resident before timing. The online scope includes queried path construction and repetition, with host transfer, label generation and training excluded from these online timings. Numerical propagation uses the same one-period reuse as QEN, so the comparison measures the online cost of producing the reusable one-period information under the specified implementation and precision criterion.

\begin{table}[t]\centering\footnotesize\setlength{\tabcolsep}{3pt}
\caption{Online query timing under a common error threshold on RTX 5090. Times are medians of five synchronized repetitions after two warmups. Each task selects its numerical resolution using per-member worst-time process-infidelity p95 on validation. Inputs and analytic reference endpoints are resident before timing.}\label{tab:onlinecost}
\begin{tabular}{rlrrr}\toprule
Batch & Query & QEN (ms) & Numerical (ms) & Ratio\\\midrule
1 & Endpoint & 3.75 & 11974 & 3195.9\\
1 & Single-period path & 13.76 & 2991 & 217.3\\
1 & 24-period grid & 15.31 & 11942 & 780.1\\
256 & Endpoint & 5.34 & 12458 & 2334.9\\
256 & Single-period path & 158.88 & 3131 & 19.7\\
256 & 24-period grid & 170.75 & 12480 & 73.1\\
\bottomrule\end{tabular}\end{table}

\begin{table}[t]\centering\small
\caption{Validation precision used to select numerical resolution for the query benchmark. Each error is process infidelity maximized over the task's queried times for each member, followed by parameter p95. Both QEN and numerical propagation are compared with the same high-resolution reference. The numerical method is fourth-order commutator-free propagation.}\label{tab:matchingprecision}
\begin{tabular}{lrrr}\toprule
Query & QEN threshold & Numerical error & Steps\\\midrule
Endpoint & $1.34\times10^{-5}$ & $1.72\times10^{-6}$ & 2048\\
65-node path & $1.52\times10^{-3}$ & $3.63\times10^{-5}$ & 512\\
24-period grid & $2.36\times10^{-3}$ & $5.28\times10^{-4}$ & 2048\\
\bottomrule\end{tabular}\end{table}

\clearpage
%

\end{document}